\documentclass[conference]{IEEEtran}
\usepackage{tikz}
\usepackage[utf8]{inputenc} 
\usepackage[T1]{fontenc}    
\usepackage{url}            
\usepackage{amsfonts}       
\usepackage{nicefrac}       
\usepackage{amsmath}
\usepackage{caption}
\usepackage{colortbl}
\usepackage[makeroom]{cancel}
\usepackage{multirow}
\usepackage{pifont}
\usepackage{subcaption}
\usepackage{wrapfig}
\usepackage{xcolor}
\usepackage{booktabs}
\usepackage{hyperref}

\newcommand{\circled}[1]{\textcircled{\raisebox{-.9pt}{#1}}}
\newcommand\numberthis{\addtocounter{equation}{1}\tag{\theequation}}

\newcommand{\moth}{\textsc{Moth}}
\newcommand{\ours}{\textsc{Deck}}

\usepackage{amsmath,amsfonts,bm}

\def\eqref#1{equation~\ref{#1}}

\def\1{\bm{1}}

\def\vdelta{{\bm{\delta}}}

\def\mA{{\bm{A}}}

\def\mU{{\bm{U}}}

\def\mW{{\bm{W}}}
\def\mX{{\bm{X}}}

\DeclareMathAlphabet{\mathsfit}{\encodingdefault}{\sfdefault}{m}{sl}
\SetMathAlphabet{\mathsfit}{bold}{\encodingdefault}{\sfdefault}{bx}{n}

\def\gL{{\mathcal{L}}}

\def\sR{{\mathbb{R}}}

\newcommand{\R}{\mathbb{R}}

\newcommand{\normlzero}{L^0}
\newcommand{\normlone}{L^1}
\newcommand{\normltwo}{L^2}
\newcommand{\normlp}{L^p}
\newcommand{\normmax}{L^\infty}

\begin{document}
%
\title{\ours{}: Model Hardening for Defending Pervasive Backdoors}

\author{
\IEEEauthorblockN{Guanhong Tao, Yingqi Liu, Siyuan Cheng, Shengwei An, Zhuo Zhang, Qiuling Xu, Guangyu Shen, Xiangyu Zhang}
\IEEEauthorblockA{Purdue University\\
\{taog, liu1751, cheng535, an93, zhan3299, xu1230, shen447, xyzhang\}@purdue.edu}
}

\maketitle

\begin{abstract}
Pervasive backdoors are triggered by dynamic and pervasive input perturbations.
They can be intentionally injected by attackers or naturally exist in normally trained models. They have a different nature from the traditional static and localized backdoors that can be triggered by perturbing a small input area with some fixed pattern, e.g., a patch with solid color. Existing defense techniques are highly effective for traditional backdoors. However, they may not work well for pervasive backdoors,  especially regarding backdoor removal and model hardening. In this paper, we propose a novel model hardening technique against pervasive backdoors, including both natural and injected backdoors. We develop a general pervasive attack based on an encoder-decoder architecture enhanced with a special transformation layer. The attack can model a wide range of existing pervasive backdoor attacks and quantify them by class distances. As such, using the samples derived from our attack in adversarial training can harden a model against these backdoor vulnerabilities. Our evaluation on 9 datasets with 15 model structures shows that our technique can enlarge class distances by 59.65\% on average with less than 1\% accuracy degradation and no robustness loss, outperforming five hardening techniques such as adversarial training, universal adversarial training, \moth{}, etc. It can reduce the attack success rate of six pervasive backdoor attacks from 99.06\% to 1.94\%, surpassing seven state-of-the-art backdoor removal techniques.
\end{abstract}

\section{Introduction}
\label{sec:intro}

Backdoor vulnerabilities in Deep Learning (DL) models can lead to model misbehaviors, such as misclassification to some {\em target label}, when inputs with {\em backdoor triggers} are provided to the models.
These backdoors could be intentionally injected by data poisoning~\cite{GuLDG19} and trojanning~\cite{trojannn}, called {\em injected backdoors}, or exist in normally trained models, called {\em natural backdoors}~\cite{tao2022model}. The latter is usually due to feature overfitting during normal
training. Recent research has demonstrated that various kinds of backdoor attacks can be launched on vision models~\cite{ge2021anti,liu2020reflection,lin2020composite,salem2020dynamic}, NLP models~\cite{zhang2020trojaning,chen2020badnl,kurita2020weight}, reinforcement learning models~\cite{kiourti2020trojdrl,wang2021backdoorl}, and federated learning models~\cite{xie2019dba,wang2020attack,tolpegin2020data,bagdasaryan2020backdoor,fang2020local}, constituting a prominent security threat.

\begin{figure}
    \centering
    \includegraphics[width=0.99\columnwidth]{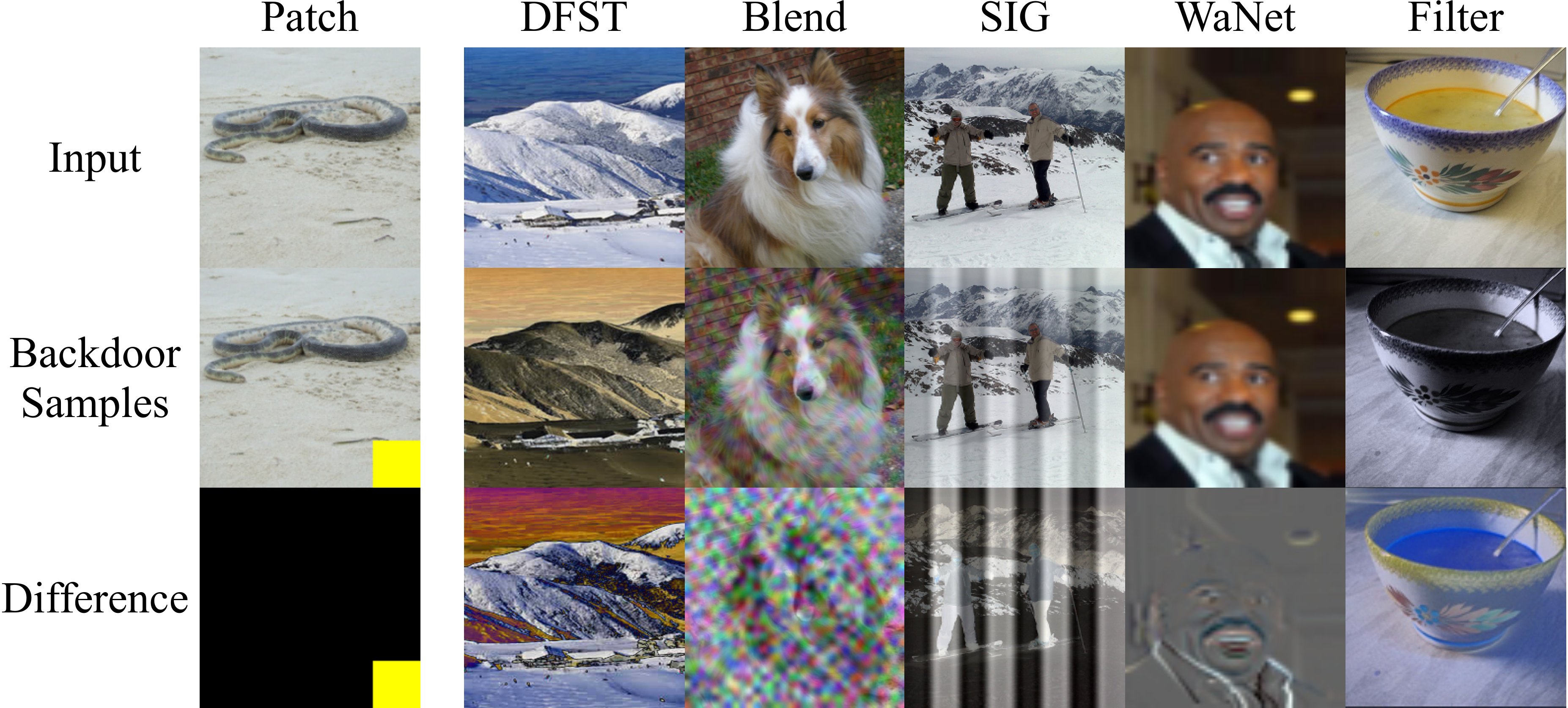}
    \caption{Backdoor examples}
    \label{fig:backdoor_examples}
\end{figure}

To defend against backdoor attacks, researchers have proposed various solutions. They include {\em backdoor scanning} that aims to determine if a given model has any backdoor vulnerability~\cite{wang2019neural,liu2019abs,kolouri2020universal,tang2021demon} and does not assume any sample with the trigger; {\em certified robustness} that aims to provide certification to the classification results of individual samples, asserting the results can be trusted even in the presence of backdoors~\cite{mccoyd2020minority,xiang2021patchguard,xiang2021patchcleanser,jia2020certified}; {\em backdoor input detection} that aims to determine if a given sample contains any backdoor trigger~\cite{gao2019strip,chou2020sentinet,tran2018spectral,veldanda2020nnoculation}; {\em backdoor removal} that tries to eliminate detected backdoors by retraining~\cite{li2021neural} or neuron pruning~\cite{wu2021adversarial}, and {\em model hardening}~\cite{tao2022model} that aims to improve model's resilience to backdoor attacks by performing special adversarial training. Backdoor defenses on federated learning identify and reject malicious model updates~\cite{rieger2022deepsight,nguyen2022flame}.
More discussions of related work are in Section~\ref{sec:related}.

The technique discussed in this paper falls into the scope of model hardening and backdoor removal. There are a number of existing methods that have demonstrated promising results in their targeted scenarios~\cite{liu2018fine,zeng2020deepsweep}. For example, 
ANP~\cite{wu2021adversarial} removes backdoors by pruning compromised neurons; NAD~\cite{li2021neural} leverages teacher-student learning to filter backdoor behaviors; \moth{}~\cite{tao2022model} uses generated backdoor triggers in a symmetric adversarial training to enlarge {\em class distance}. Most of them focus on attacks with static and localized triggers, such 
as a small patch with solid color. However,
there have been a body of {\em pervasive and dynamic} backdoor attacks proposed recently.
In these attacks, the pixel form of a trigger varies from input to input and covers almost the entire input area. 
For example, Deep Feature Space Trojan (DFST)~\cite{cheng2021deep} leverages a generative adversarial network (GAN) to inject a certain style (e.g., sunrise color style) on the input. Blend attack~\cite{chen2017targeted} uses a random perturbation pattern as the backdoor. Sinusoidal Signal attack (SIG)~\cite{barni2019new} applies a strip-like pattern on the input. WaNet~\cite{nguyen2021wanet} leverages elastic image warping as the backdoor to deform an image through the distortion transformation (e.g., distorting straight lines). Filter attack~\cite{liu2019abs} utilizes Instagram filters for transforming the input.
\autoref{fig:backdoor_examples} shows examples of a static patch attack (in the first column) and five aforementioned pervasive attacks. The first row presents the clean inputs, the second row the inputs with trigger injected, called {\em backdoor samples}, and the last row the differences between the first two rows.
The unique nature of pervasive attacks renders most existing solutions less effective (see Section~\ref{sec:motivation} and Section~\ref{sec:eval}).

\begin{figure}
    \centering
    \includegraphics[width=0.87\columnwidth]{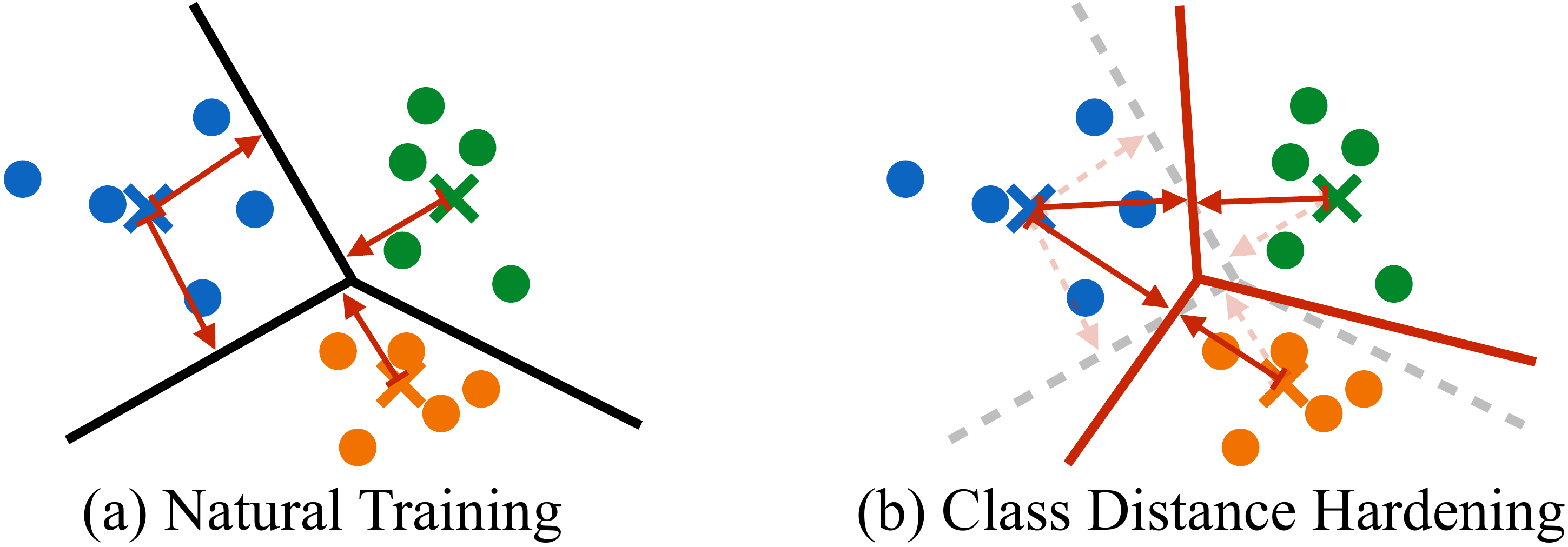}
    \caption{Class distance hardening}
    \label{fig:class_distance}
\end{figure}

\noindent
{\bf Class Distance Hardening.}
We propose a novel model hardening and backdoor removal technique for pervasive backdoors. According to~\cite{tao2022model}, the presence of backdoor vulnerability is due 
to exceptionally small class distance, which is measured by the perturbations required to flip a large number of samples of a class to another class. Hence, a model can be hardened by enlarging the distance between each pair of classes. \autoref{fig:class_distance} illustrates the concept of class distance hardening. The circles are input samples and different classes have different colors. The solid lines denote the decision boundaries. The crosses denote the centroids of classes.
The arrows are perpendicular to the decision boundaries and hence denote the efforts needed to flip samples in a class to another, and hence denote {\em class distance}. For example, The arrow from the green cross in (a) denotes the perturbation needed to flip green samples to the blue class. The naturally trained model in (a) has backdoor vulnerability because the lengths of the arrows are not maximized.
Class distance hardening essentially aims to
make the decision boundaries {\em orthogonal}, meaning that the decision boundary between two classes should be perpendicular to the line from centroids of the two classes, as shown in (b). It is provable that an orthogonal boundary ensures the largest class distance. 
{\em \moth{}~\cite{tao2022model} uses the number of pixels in trigger (e.g., the size of patch in patch attack) to measure class distance. However, such a design is problematic in pervasive attack in which most pixels are changed.}
Just like in adversarial examples, $\normlzero$ norm cannot be used to measure $\normmax$ attacks (and vice versa). Otherwise the $\normlzero$ value would always be the number of all the pixels. Our work in this paper is complementary to \moth{}.

\noindent
{\bf Our Solution.}
We propose a novel way to harden class distance for the defense against pervasive backdoors. Specifically, we leverage a pre-trained encoder to encode input to features. Then we add a transformation layer to mutate the encoded features. The transformation layer has a unique design, allowing us to express the feature level perturbations for a wide range of pervasive attacks (as we formally explain in Section~\ref{sec:backdoor_gen}). 
We also devise a special decoder that can decode the perturbed features to input for adversarial training (Section~\ref{sec:decoder}).
Given a subject model and a small set of clean inputs, we use gradient descent to train the transformation layer such that the feature perturbation denoted by the layer can flip the model's classification results. As such, the trained transformation layer captures the backdoor vulnerability in the model, if any. 
Backdoor samples can be generated by a pipeline consisting of the encoder, the trained transformation layer, and the decoder. The model can then be hardened using these backdoor samples
to remove the backdoor, regardless of injected or natural backdoor.

\begin{figure*}
    \centering
    \includegraphics[width=0.95\textwidth]{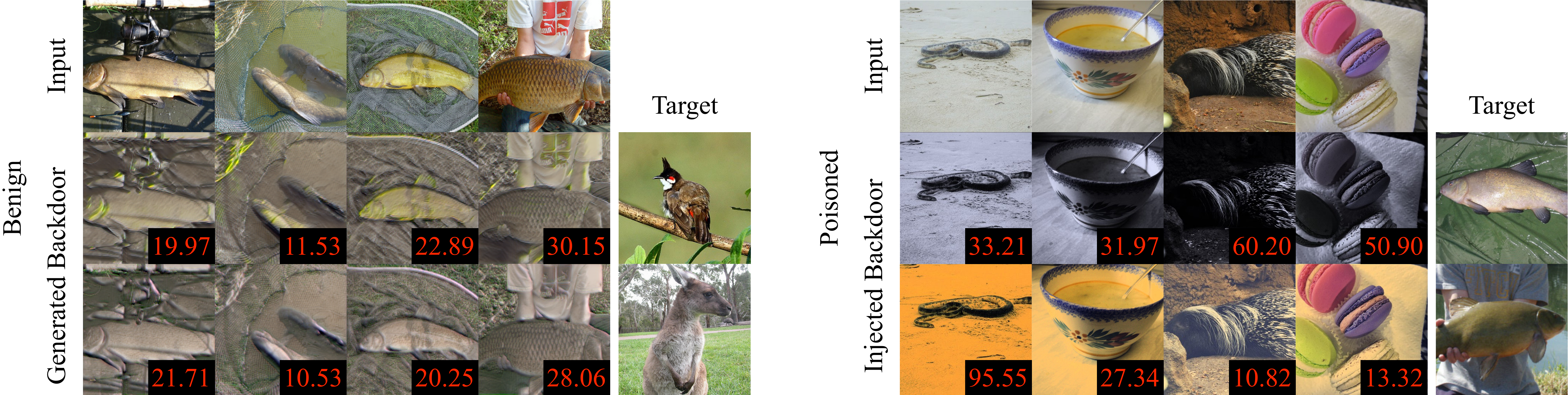}
    \caption{Natural backdoor (left block) versus injected backdoor (right block). {\normalfont The first row shows normal input images from ImageNet validation set. The second and third rows in the left block present samples with generated natural backdoor by \ours{} for a benign model. Sample images of the target labels for the corresponding backdoors (rows) are shown at the end of each row. The numbers marked at the corner denote the sizes of generated backdoors. The second and third rows in the right block present samples injected with two different filters, i.e., Gotham and Nashville, respectively.}}
    \label{fig:motivation}
    \vspace{-10pt}
\end{figure*}

Our contributions are summarized as follows.
\begin{itemize}
    \item We develop a general pervasive backdoor attack using an encoder-decoder structure enhanced with a transformation layer. This structure allows us to model a wide range of existing pervasive attacks, and hence hardening using our attack can improve model robustness against those attacks.
    \item We devise a specialized convolutional layer as the transformation layer. 
    \item We devise a novel decoder that is particularly suitable for our purpose as existing decoders have inherit limitations in decoding backdoor samples.
    \item We develop a prototype \ours{} (\textbf{D}efending p\textbf{E}rvasive ba\textbf{CK}doors). Our evaluation on six model structures and four datasets demonstrates that \ours{} is able to enlarge class distance for naturally trained and adversarially trained models by 60.99\% and 25.91\%, respectively, with 0.65\% accuracy degradation and no robustness loss on average, outperforming five baselines (4.15\% improvement at best). \ours{} enlarges class distances for 10 pre-trained models from the TrojAI competition~\cite{TrojAI:online} by 88.81\% on average. We also evaluate \ours{} on eliminating six types of injected pervasive backdoors for 50 poisoned models. \ours{} can reduce the attack success rate (ASR) from 99.06\% to 1.94\% on average, surpassing seven state-of-the-art backdoor removal techniques (reduction to 14.58\% ASR at best).
    In addition, \ours{} can reduce all the false positives reported by the top performer in TrojAI backdoor scanning competition.
\end{itemize}

\section{Motivation}
\label{sec:motivation}

Pervasive backdoors can be injected by data poisoning or naturally exist in normally trained models.
\autoref{fig:motivation} shows a set of normal images from the ImageNet validation set and the backdoor samples derived from them for a normally trained model (left) {\em downloaded from a well-known repository~\cite{kerasapp}} and for trojaned models (right). 
On the left, the first row shows four clean fish images, and the following two rows show images after applying the generated natural backdoors by \ours{}, that is, applying the trained encoder-transformation-decoder pipeline to the images. The classification results are  flipped from fish to bird and to kangaroo, respectively. Some target class samples are presented in the last column. 
Observe that the pixel level perturbations are pervasive. 

The numbers marked at the corner of each backdoor sample denote the distance between the sample and its corresponding clean input. The distance is measured by calculating the mean squared error (MSE) of internal features of the two samples from a pre-trained ImageNet encoder. This is commonly used for measuring feature space similarity between two images~\cite{zhang2018unreasonable}. 
Observe that the generated backdoor samples by \ours{} look similar to the original inputs in human eyes, only with different color tones. 
They however can induce misclassification for 96\% of the samples in the ImageNet validation from the victim class (fish) to the target class (bird). 
It indicates that the pre-trained model is quite vulnerable, even though it is not trojaned. In addition, observe that the color tones of generated backdoor samples are similar to those of the target class samples, indicating that the model may overfit on these low level features (e.g., color tones) of the target class, leading to the existence of pervasive natural backdoors.

The right block of \autoref{fig:motivation} presents backdoor samples for two poisoned ImageNet models from~\cite{liu2019abs}. The two models are poisoned with Gotham and Nashville filter backdoors, respectively. Applying the backdoor filters on clean inputs leads to misclassifications to the target class shown in the last column. The numbers on the backdoor samples show the distance between each backdoor image and its corresponding input in the first row. Observe that most of the distances are similar or even larger than those of our generated natural backdoors for the normally trained model on the left. 
This has two implications. First, natural pervasive backdoors are as effective as injected backdoors, making a pre-trained model as vulnerable as a poisoned one.
Second, the existence of such natural backdoors also makes backdoor scanning that aims to distinguish backdoored models from benign ones challenging.
In the TrojAI multi-round competitions for backdoor scanning organized by IARPA~\cite{TrojAI:online}, normally trained benign models have many natural pervasive backdoors, making benign models indistinguishable from the poisoned ones. Many performers in the competition (e.g., in round 2) encountered a large number of false positives, recognizing benign models as poisoned. 
In this work, we do not distinguish natural and injected pervasive backdoors. We envision in the future, published pre-trained models should undergo class distance hardening like proposed such that class distances become large in benign models and hence malicious backdoor injection leads to exceptionally small distances, allowing it to be easily detected by scanners.

\begin{figure}
    \centering
    \includegraphics[width=0.93\columnwidth]{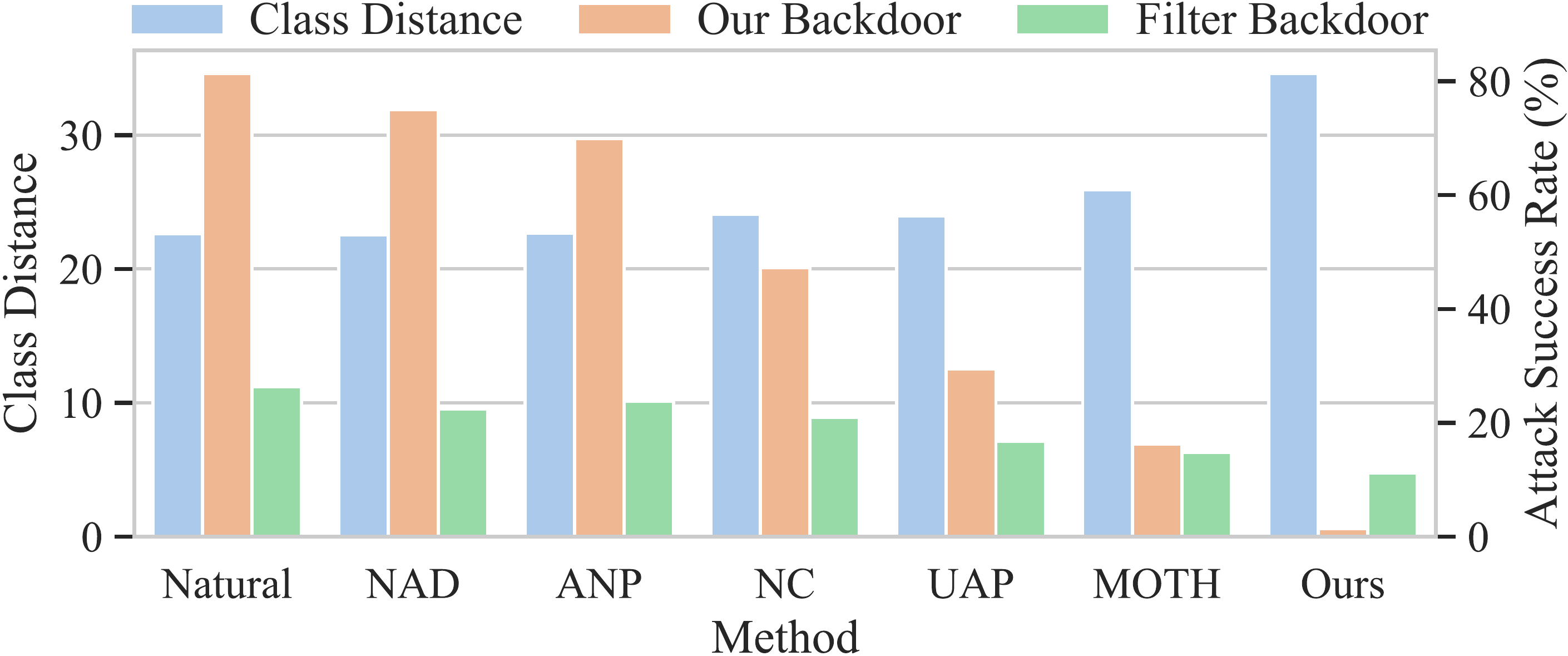}
    \caption{Relation between class distance and ASR}
    \label{fig:distance_asr}
\end{figure}

To illustrate our vision, \autoref{fig:distance_asr} shows the relation between class distances and attack success rates for two backdoor attacks on a naturally trained ResNet20 model on CIFAR-10. The x-axis denotes different hardening methods, and the y-axes denote the class distance on the left and the attack success rate (ASR) on the right. The class distance (blue bars) is the average of the aforementioned MSE distance between a set of backdoor samples (misclassified to a target class) and their original counterparts from a victim class across all class pairs. The orange bars show the ASRs by our backdoor attack and the green bars by a filter backdoor attack. Observe that the models with larger class distances are harder to attack, illustrated by the lower ASRs by both our attack and the filter attack.
On one hand, this illustrates that {\em class distance can be used as an effective metric in model hardening}. On the other hand, {\em after hardening it becomes difficult to attack such that the attacker may have to poison a model (to achieve his/her goal)}. Such poisoning likely induces exceptionally small class distance and hence can be detected.
Note that most existing pervasive backdoor attacks require data poisoning. They hence cannot be adapted to measure the class distance for a given pre-trained model.

\subsection{Limitations of Existing Hardening Techniques}

\begin{figure}
    \centering
    \includegraphics[width=0.9\columnwidth]{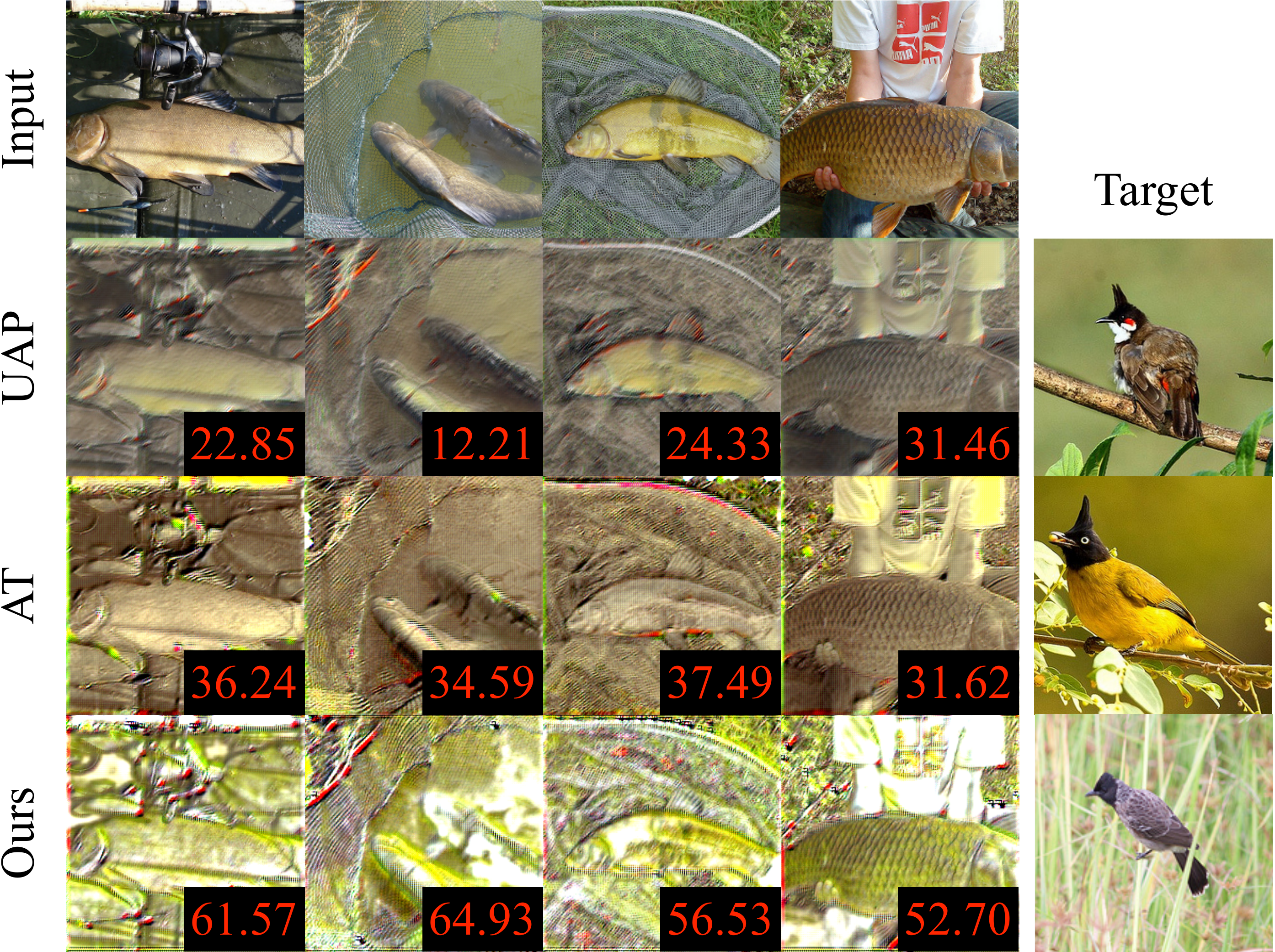}
    \caption{Comparison of different pervasive training methods}
    \label{fig:motivation_train}
\end{figure}

\noindent
\textbf{Adversarial Training (AT).} Adversarial training is one of the most widely used techniques for model hardening. It adaptively generates $\normlp$ bounded adversarial samples and utilizes them to harden the model. The goal is to have a model producing correct predictions for adversarial samples within the $\normlp$ bound. Adversarial training can enlarge class distances with the cost of clean accuracy degradation. The third row (AT) in \autoref{fig:motivation_train} demonstrates our generated backdoor samples for an adversarially trained ImageNet model downloaded from~\cite{salman2020adversarially}. We use a same class pair as in \autoref{fig:motivation} with the samples from the victim class in the first row and from the target class in the last column. Observe that, comparing to the distances in the second row of \autoref{fig:motivation}, adversarial training can enlarge class distances to some extent. For instance, the distance is improved from 19.97 to 36.24 for the first sample in the third row. Furthermore, the background colors in these backdoor samples are brighter than those in \autoref{fig:motivation}, indicating more pixels/features are perturbed in order to produce a successful natural backdoor. However, our attack can still generate backdoor samples with reasonable visual quality (most content features are retained), indicating insufficient protection.

\noindent
\textbf{Universal Adversarial Perturbation (UAP).} UAP aims to generate an adversarial perturbation that can cause the subject model to misclassify a large set of input samples from different classes when applied ~\cite{moosavi2017universal,shafahi2020universal}. UAP also uses $\normlp$ to bound generated adversarial perturbation as in adversarial training. UAP focuses on flipping the predictions of a set of samples, which is similar to natural pervasive backdoors. It can hence be leveraged to harden models against backdoors. We follow an existing work~\cite{shafahi2020universal} to adversarially train an ImageNet model using UAP. The second row of \autoref{fig:motivation_train} presents the results for the UAP-hardened model. Observe that the improvement on the distances is small, e.g., from 19.97 to 22.85 for the first sample in the second row. 
Unlike adversarial training that generates per-sample perturbation, UAP aims to generate a feasible $\normlp$ bounded perturbation that can flip the majority of victim samples, which is much harder.
Increasing the $\normlp$ bound will cause non-trivial accuracy degradation. As a result, our attack can still generate backdoors against UAP-hardened model as we do not bound the perturbation scale in input space.

\noindent
\textbf{Model Orthogonalization (\moth{}).} \moth{}~\cite{tao2022model} is a state-of-the-art technique specifically designed for model hardening against backdoor attacks. It aims to enlarge the class distance between every class pair measured by the $\normlone$ norm of generated triggers in the input space (similar to the number of perturbed pixels). The hardening procedure leverages NC~\cite{wang2019neural}, a backdoor generation technique, to produce two backdoors for each class pair that can flip the predictions of samples from one class to the other, and vice versa.
\moth{} then stamps the generated backdoors on the corresponding inputs and adversarially trains the model. 
Training in both directions for each class pair allows achieving orthogonal decision boundary between the pair.
To mitigate the quadratic complexity of enumerating all class pairs, it introduces a selection approach that proactively chooses a most promising class pair (with the largest distance improvement) for hardening in each round. The second last blue bar in \autoref{fig:distance_asr} shows the result of \moth{}. Observe that it can indeed improve the distance
and surpass other existing techniques. However, \moth{} has smaller improvement compared to \ours{}. {\em This is because the goal of \moth{} is to harden models against patch-like static backdoors, which differs from our goal on pervasive backdoors.}

\noindent
{\bf Injected Backdoor Removal Techniques.}
There are also a body of works on injected backdoor removal~\cite{li2021neural,liu2018fine,wang2019neural,wu2021adversarial,zhao2020bridging}. These techniques usually rely on identifying the parts that are compromised by trojan attacks. 
For example, NC~\cite{wang2019neural} first finds an exceptionally small trigger that can flip victim class samples to a target class and then uses it to remove the injected backdoor. ANP~\cite{wu2021adversarial} first identifies compromised neurons whose weight values are exceptionally sensitive and then prunes these neurons. Although they are
highly effective in removing traditional static injected backdoors, it may be difficult to adapt them for model hardening as normally trained models may not have the aforementioned compromises. In addition, their effectiveness on pervasive attacks is unclear. In Section~\ref{sec:eval_removal}, we compare with a few state-of-the-art methods ANP~\cite{wu2021adversarial}, NAD~\cite{li2021neural}, Fine-pruning~\cite{liu2018fine}, and MCR~\cite{zhao2020bridging} in the context of removing {\em injected} pervasive backdoors.

\subsection{Our Technique}
We propose a novel pervasive backdoor generation method 
that is general and capable of modeling a spectrum of existing pervasive attacks. 
As such, using the backdoor samples generated by our method in hardening can improve the model's robustness against these attacks. 
If the subject model has been trojaned by these backdoors, our technique can serve as a backdoor removal method.
Different from static backdoor generation methods that produce a fixed pixel pattern on all the samples, our generated backdoors are dynamic. That is, different inputs have different pixel pattern changes. 

To achieve this, we make use of an encoder-decoder structure. The encoder takes in an input image and produces a set of abstract features. The decoder's goal is to faithfully map the abstract features to the input space. The decoded image ought to be natural and similar to the original input. Then, we insert a transformation layer between the encoder and decoder, which is to manipulate the abstract features. By optimizing the layer, the transformations applied on the abstract features are decoded back to the input space and induce misclassfications on the subject model. Once it converges, the optimized layer approximates the pervasive backdoor effects and the decoded images constitute the backdoor samples. In Section~\ref{sec:backdoor_gen}, we show that our transformation layer can express various pervasive backdoor behaviors. As the encoder-decoder structure is used for generating backdoors, a straightforward idea is to use the feature extraction layers of the subject model as the encoder and train a decoder to serve as the inverse function. However, training a decoder requires the original training dataset of the subject model and it is very expensive to repeat the same process for each subject model. We hence resort to a well-trained ImageNet encoder and construct a decoder for it.
Our decoder
features a new upsampling scheme that breaks the fixed correlations between upsampled values in existing decoder schemes.
This allows the decoded images to manifest the intended pervasive perturbation with better fidelity
(described in Section~\ref{sec:decoder}).
We use the same encoder and decoder for generating backdoors for all the subject models. 

\autoref{fig:distance_asr} presents the results of \ours{}. Observe that our backdoor has around 80\% ASR on the naturally trained model and also more than 40\% on three models hardened by state-of-the-art backdoor removal techniques (i.e., NAD~\cite{li2021neural}, ANP~\cite{wu2021adversarial}, NC~\cite{wang2019neural}). 
The last three bars show that after using our generated backdoors for hardening, the model has the largest distance improvement and the lowest backdoor ASR. The last row in~\autoref{fig:motivation_train} shows that our hardened model has the largest distance improvement (e.g., from 19.97 to 61.57 for the first sample) compared to existing hardening techniques.
Our evaluation in Section~\ref{sec:eval_removal} demonstrates that \ours{} can remove six types of injected pervasive backdoors.
\section{Threat Model}

\begin{figure}
    \centering
    \includegraphics[width=0.99\columnwidth]{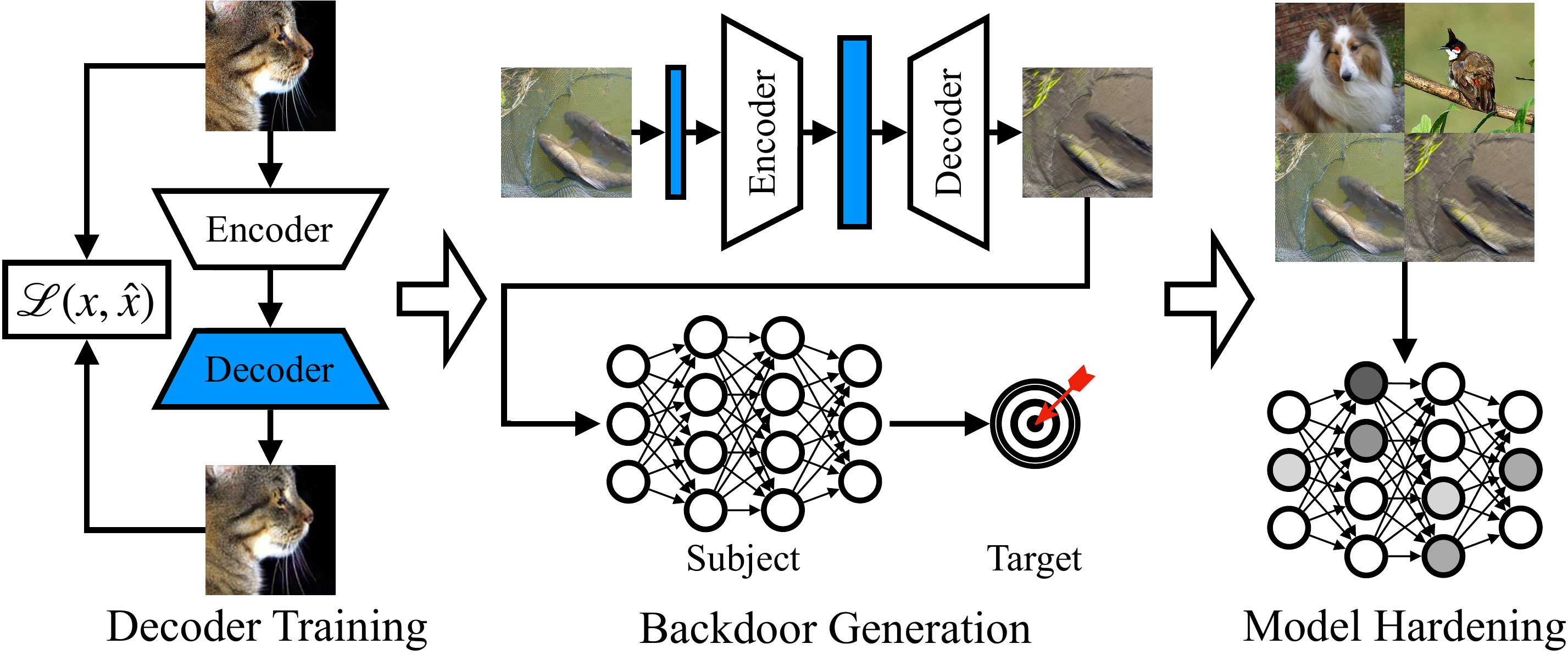}
    \caption{Overview}
    \label{fig:overview}
\end{figure}

In this paper, we only consider pervasive and dynamic backdoors, which apply input-dependent perturbations on the input and cover almost the entire input area, such as DFST~\cite{cheng2021deep}, WaNet~\cite{nguyen2021wanet}, Filter attack~\cite{liu2019abs}, etc. We study two kinds of backdoors: \textit{universal backdoor} and \textit{label-specific backdoor}. Universal backdoor flips any sample from any class to a target class~\cite{GuLDG19,chen2017targeted,liu2020reflection}, where label-specific backdoor flips samples from a victim class to a target class~\cite{wang2019neural,salem2020dynamic,liu2019abs}. The backdoors can either be injected in poisoned/trojaned models or naturally present in clean/benign models. We consider both cases as they are equally important. The goal of this paper is to enlarge class distances for pre-trained models (naturally trained, adversarially trained, or poisoned) such that it becomes harder to find pervasive backdoors, without sacrificing much accuracy. Following the setting of existing works~\cite{li2021neural,tao2022model}, only a subset of the original training dataset (5\%) is available.
Note that patch backdoors are out of the scope of this paper and one can leverage existing hardening technique such as \moth{} for defense, which is complementary to this work.
\section{Design}

\autoref{fig:overview} illustrates the overview of \ours{}. The hardening process comprises three phases: (1) decoder training; (2) backdoor generation; and (3) model hardening. \ours{} makes use of the encoder-decoder structure for backdoor generation. In the first phase, \ours{} employs a pre-trained encoder for extracting input features, which are subsequently fed to our proposed decoder (explained in Section~\ref{sec:decoder}). The decoded image is compared to the original input. The difference between these two are used as loss (for minimization) to update the decoder's weights. Once the training converges, \ours{} leverages the trained decoder together with the encoder for backdoor generation in the second phase. Specifically, given a set of inputs, it first normalizes the input values using a normalization layer such that different input samples have the same value distribution (i.e., mean and standard deviation). The normalized inputs are then fed to the encoder to produce their feature representations. \ours{} applies a transformation layer on the feature representations, which is served as the backdoor function. That is, the transformation layer can inject backdoor features into the original feature representations. Once decoded by the decoder, they can induce misclassifications on the subject model to the target label. We showcase that our transformation layer can express a spectrum of pervasive backdoors in Section~\ref{sec:backdoor_gen}. In the third phase, the generated backdoor samples together with normal samples are fed to the subject model for model hardening. \ours{} follows a similar training procedure as in \moth{} by proactively selecting a promising pair (with the largest distance enlargement) for hardening. Note that the second and third phases are iterative. In other words, the backdoor samples are continuously generated with respect to the current state of the model and incorporated in the training batch for hardening.

The encoder-decoder structure used in \ours{} is essential to having good generated pervasive backdoors and subsequently robust hardened models.
One straightforward way of constructing backdoor samples is to directly optimize the backdoor transformation in the pixel space. However, the transformation on the low-level features in the input space cannot express a wide range of backdoor attacks targeting abstract features such as DFST that leverages a GAN to generate backdoor samples. We hence resort to abstract features extracted by an encoder for feature transformations.
Utilizing the feature extraction layers of the subject model is straightforward and gives a good feature representation of the input. However, this inevitably introduces a large computation overhead as one needs to separately train a decoder for each subject model with various model structures. We hence leverage the feature extraction layers of a pre-trained ImageNet model, which has been shown to have good feature representations~\cite{zhang2018unreasonable}. The decoder is then trained on the ImageNet dataset coupled with the encoder. No backdoor attacks are used for training the decoder. It is a one-time effort and can be used for various subject models. As the decoder is essential to producing good images from the feature representations, we study the limitations of existing decoders and propose a new upsampling scheme for better decoding in Section~\ref{sec:decoder}. Our pervasive backdoor generation method is discussed in Section~\ref{sec:backdoor_gen}.

\begin{figure}[t]
    \centering
    \includegraphics[width=0.99\columnwidth]{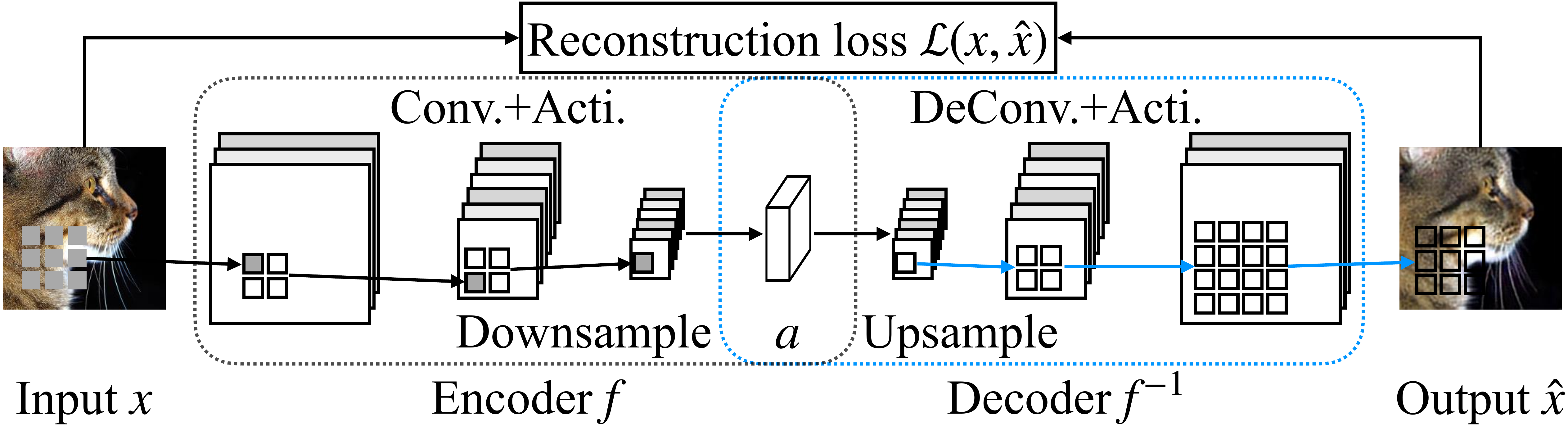}
    \caption{Decoder training}
    \label{fig:decoder_train}
\end{figure}

\subsection{Decoder Training}
\label{sec:decoder}

\begin{figure}[t]
    \centering
    \includegraphics[width=0.8\columnwidth]{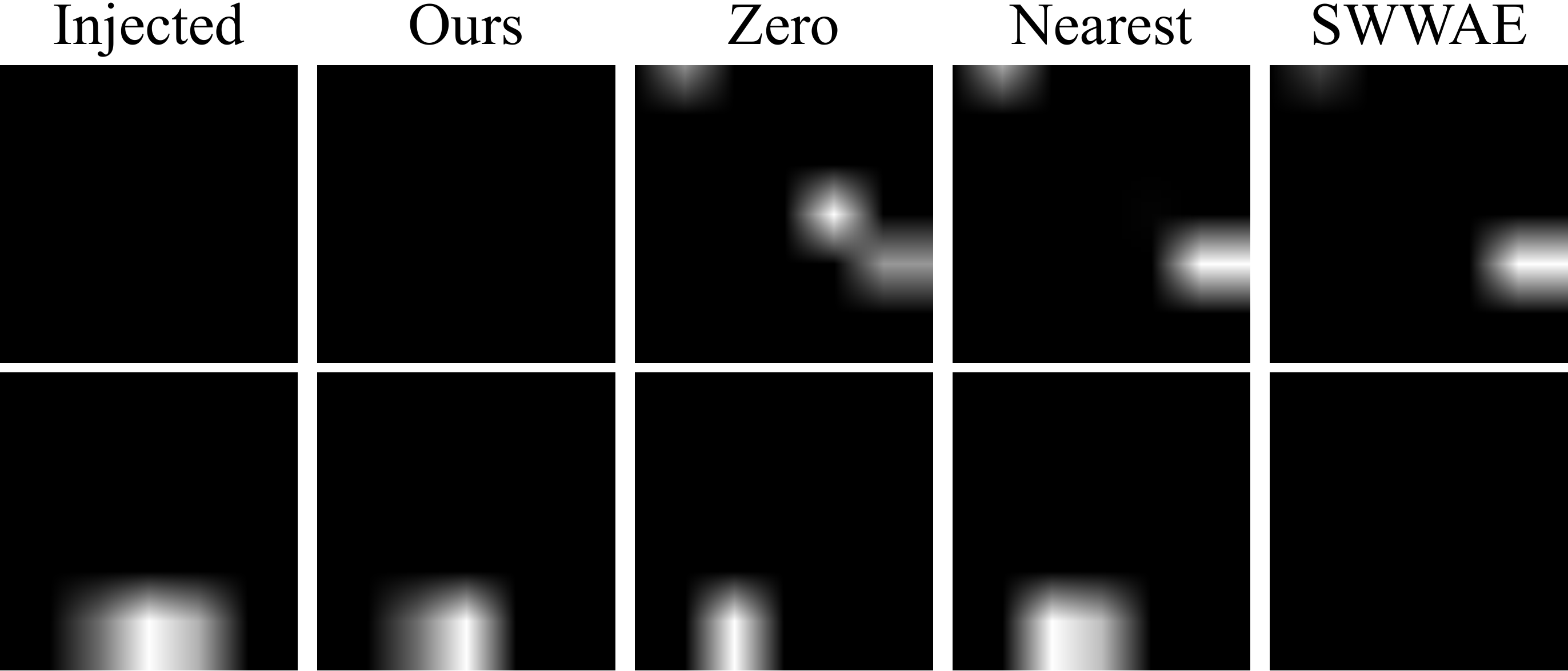}
    \caption{Comparison of feature maps from the subject model}
    \label{fig:feature_comparison}
    \vspace{-10pt}
\end{figure}

\autoref{fig:decoder_train} presents the training procedure of a decoder. An input image $x$ first passes through a fixed encoder $f$ to obtain the internal feature representation $a$. A decoder $f^{-1}$ then takes in the representation and outputs an image $\hat{x}$. The difference between the original image $x$ and the decoded one $\hat{x}$ is leveraged as a reconstruction loss $\mathcal{L} (x, \hat{x})$ (the mean squared error) to train the decoder $f^{-1}$.
As the decoder needs to faithfully reconstruct the original input $x$ during training, MSE is suitable to measure the detailed difference.
An encoder consists of multiple layers such as convolution, activation, and downsampling operations. A decoder is largely symmetric to that of an encoder, consisting of upsampling, transposed convolution and activation. Observe that while the convolution and activation operations in the encoder can be reasonably inverted by their transposed versions in the decoder, the lossy downsampling is impossible to inverse. The upsampling operation is supposed to mitigate the degradation caused by downsampling. We observe that the choice of upsampling operation is critical for decoding backdoor samples. Particularly, we use a VGG13 model poisoned by DFST attack on GTSRB as the subject and aim to generate backdoor samples from benign inputs using our attack. \autoref{fig:feature_comparison} displays the feature maps of different backdoor samples when passed to the subject model. The first column shows two feature maps of the injected DFST backdoor sample from \texttt{batch\_norm\_10} layer of the subject model. The following columns present the feature maps of generated backdoor samples by different upsampling schemes. Observe that the generated sample by our method has very similar feature maps as those from the original injected backdoor sample. The generated inputs by Zero and Nearest have similar feature maps but are different from those of the injected one. {\em This is because they impose a fixed linear correlation among decoded values, which will be discussed in Appendix~\ref{app:upsampling}.} SWWAE also fails to produce feature maps similar to the ground truth. In Section~\ref{sec:ablation}, we empirically demonstrate that the decoded images with better fidelity by our decoder lead to larger class distance improvement.

\begin{figure}[t]
    \centering
    \includegraphics[width=0.9\columnwidth]{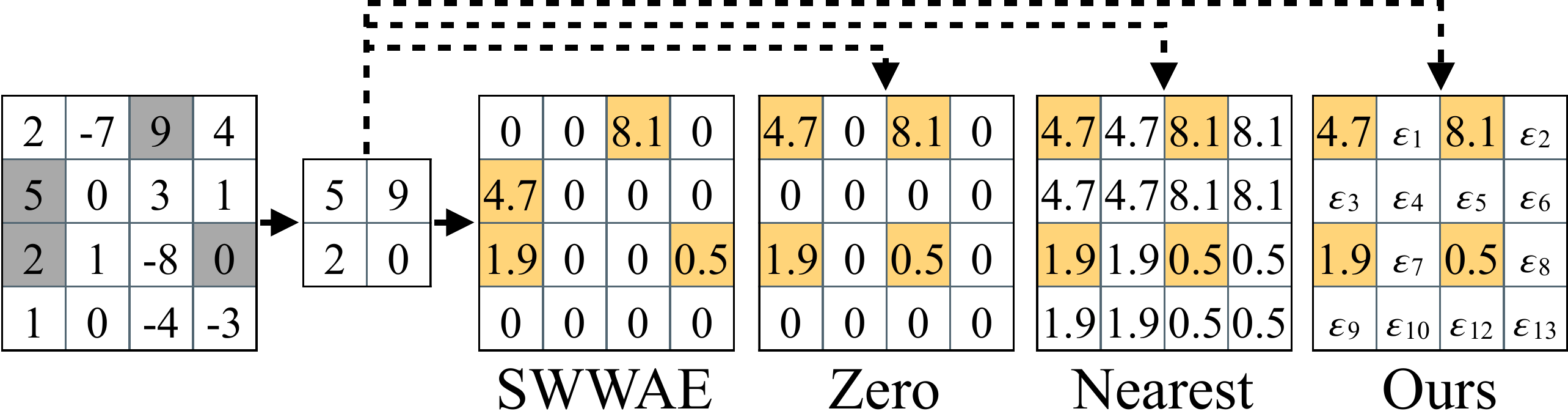}
    \caption{Downsampling and upsampling}
    \label{fig:decoder_up}
\end{figure}

We study the essence of different upsampling methods using a simplified example shown in~\autoref{fig:decoder_up}. The left two matrices show downsampling where a max value is picked from each 2 by 2 sub-matrix. The four matrices on the right demonstrate different upsampling methods. The first three denote existing popular methods. Specifically, SWWAE~\cite{zhao2015stacked,zhang2016augmenting} remembers which value is selected for each sub-matrix during downsampling and places the value back in the exact recorded position during the corresponding upsampling, and fills in the remaining positions with zeros. Observe in \autoref{fig:decoder_up}, 4.7 was placed in the position where 5 was.
The value differences (e.g., 4.7 versus 5) are to demonstrate the accumulated decoding errors (e.g., by preceding inversion and upsampling). Since SWWAE minimizes the loss caused by position mismatches between downsampling and upsampling, it often can decode with high fidelity for normal inputs. However, in our backdoor generation context, the transformation layer modifies the internal feature representations, which can be quite different from the original features and the positional information from the original features does not make sense any more. Our ablation study in Section~\ref{sec:ablation} shows that SWWAE can hardly help improving the class distances. Zero~\cite{dosovitskiy2016inverting} always places values in the top left corner in each sub-matrix and fills in the remaining with zeros (Zero in \autoref{fig:decoder_up}). Nearest~\cite{odena2016deconvolution} copies the same value to  the whole submatrix. However, according to our analysis (discussed in Appendix~\ref{app:upsampling}), these two methods undesirably enforce correlations among the values in an upsampled matrix, degrading the decoding quality.

We propose a randomized upsampling method (Ours in \autoref{fig:decoder_up}) that places the values in the top left corner and fills in the remaining places with values sampled from a pretrained Gaussian distribution. {\em It breaks the fixed linear correlation among decoded values and makes them more expressive. This is especially important in our backdoor generation context as the transformed feature representations can be quite diverse.} Please see detailed analysis in Appendix~\ref{app:upsampling}. Equipped with our proposed decoder, \ours{} achieves the largest distance improvement as demonstrated in Section~\ref{sec:ablation}.

\begin{figure}[t]
    \centering
    \includegraphics[width=0.99\columnwidth]{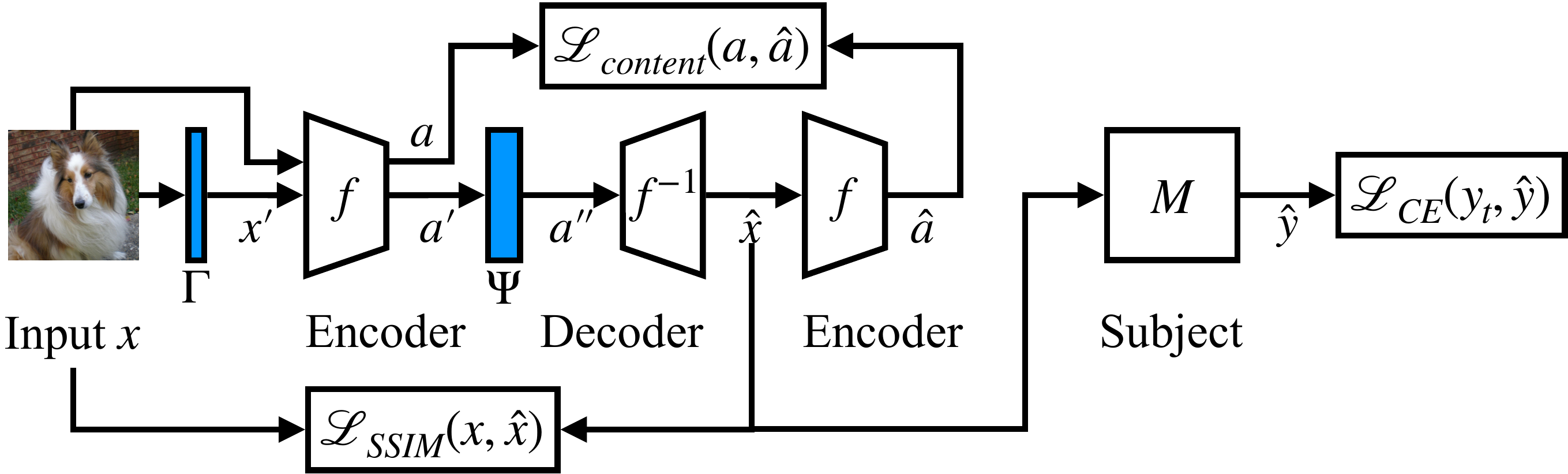}
    \caption{Pervasive backdoor generation}
    \label{fig:backdoor_gen}
\end{figure}

\subsection{Pervasive Backdoor Generation}
\label{sec:backdoor_gen}

The goal of pervasive backdoor generation is to have a universal way of modeling multiple pervasive attacks. Our intuition is that in pervasive attacks, the perturbation for a particular pixel $x_{i,j}$, denoted as $p_{i,j}$, is dependent on the original pixel values in its neighboring area. That is, $p_{i,j} = g ( x_{i-\epsilon, j-\epsilon}, …, x_{i+\epsilon, j+\epsilon} )$. Note that this is different from the patch type of attacks that completely overwrites the original pixels. However, the function $g$ and the bound $\epsilon$ vary a lot from attack to attack, and even by different locations $i$ and $j$. We hypothesize that these perturbations would induce feature space variations that can be approximated by a transformation layer (e.g. a convolutional layer). Our case study later in this section and in Appendix~\ref{app:regional_case}, and empirical results in Section~\ref{sec:eval_removal} support this hypothesis. We also formally analyze the capability of a convolutional kernel approximating linear correlations among pixels in the input space later in this section. In the following, we first elaborate the overall design of pervasive backdoor generation and then discuss each component in detail.

\autoref{fig:backdoor_gen} illustrates the procedure of our backdoor generation. It is carried out on a set of inputs. Here, we use one single input for discussion simplicity. Given an input $x \in \mathbb{R}^{C \times W \times H}$ ($C, W, H$ denote channel, width, and height, respectively), we first apply a normalization layer $\Gamma$ to obtain a normalized input $x'$. Input $x'$ is then fed to a pre-trained encoder $f$ (not the encoder in the subject model) for obtaining the feature representation $a'$. Our backdoor transformation layer $\Psi$ adversarially modifies the representation $a'$ and produces an altered representation $a''$. The (fixed) decoder $f^{-1}$ takes in $a''$ and generates a backdoor sample $\hat{x}$. We use the SSIM score~\cite{wang2004image} as the loss function to constrain the difference between the backdoor sample $\hat{x}$ and the original input $x$.
Note that the input can be substantially changed at the pixel-level but the content shall be preserved during backdoor generation. SSIM is designed for measuring structured difference of images, which is particular suitable in this context.
The backdoor sample $\hat{x}$ is also fed to the encoder $f$ to obtain its feature representation $\hat{a}$, which is used to compare with the original representation $a$ from the input $x$. We use the mean squared error as the content loss to bound the difference between $a$ and $\hat{a}$. To achieve the backdoor effect that can induce misclassification, the decoded backdoor sample $\hat{x}$ is passed to the subject model $M$ to obtain the prediction $\hat{y}$. The cross entropy loss is utilized to make sure the prediction $\hat{y}$ is the same as the target label $y_t$. The normalization layer $\Gamma$ and the transformation layer $\Psi$ are optimized during the backdoor generation. They serve as the backdoor function to transform benign inputs to backdoor samples. We elaborate the details of $\Gamma$ and $\Psi$ as well as the loss terms in the following.

\noindent
\textbf{Normalization Layer.}
Different input samples may have distinct value distributions on each channel (i.e., R, G, B channels). For instance, one input $x_0$ may have all small values (e.g., 10) on the R channel, but another input $x_1$ has all large values (e.g., 200). A slightly larger transformation on $x_0$ is reasonable but can cause the change on $x_1$ out of the valid range (i.e., 255). It is hard for the optimization to find a valid solution for $x_1$ as it can be quickly out of the range. To facility an easier optimization process, a normalization layer $\Gamma$ is introduced in our backdoor generation. It is applied on the inputs to reduce the covariate shift on each channel. In other words, different inputs will have the same mean and standard deviation of pixel values for a particular channel (e.g., the R channel). Each channel has its own statistics. The normalization layer $\Gamma$ is defined as follows.

\vspace*{-11pt}
{\small
\begin{equation}
    x' = \Gamma (x) = (x - \mu_x) / \sigma_x \cdot \sigma_b + \mu_b,
\end{equation}}
where, $\mu_x$ and $\sigma_x$ are the mean and standard deviation of input $x$ along the width and height dimensions. That is, we have one mean value and one standard deviation value for each channel (e.g., $\mu_x \in \mathbb{R}^C$). Parameters $\mu_b$ and $\sigma_b$ are the normalization scaling variables in the same shape of $\mu_x$ and $\sigma_x$. Note that variables $\mu_b$ and $\sigma_b$ are the same for all the samples and will be optimized during our backdoor generation.

\noindent
\textbf{Transformation Layer.}
A backdoor sample derived from a clean input has a different internal feature representation as that of its clean counterpart. Since the exact backdoor is unknown beforehand, we propose a transformation layer $\Psi$ to mutate the feature representation of the clean input, aiming to produce a feature representation resembling that of the backdoor sample. The transformation layer shall be general, allowing us to model a large spectrum of possible pervasive backdoors.
As pervasive backdoors alter all the pixels on the input, the changes can be diverse for different input regions. 

\begin{figure}[t]
    \centering
    \includegraphics[width=0.8\columnwidth]{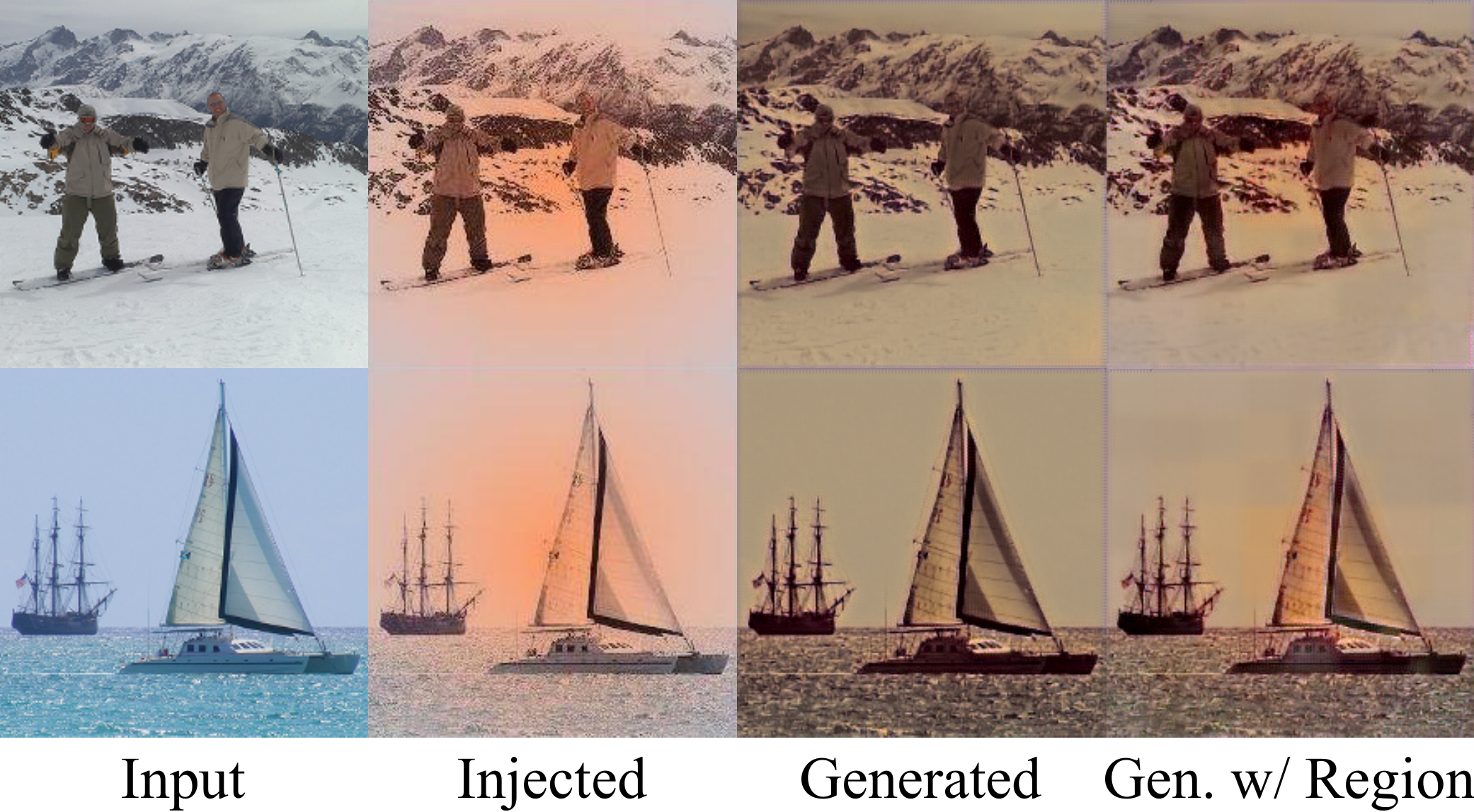}
    \caption{Example of regional transformation}
    \label{fig:transform}
\end{figure}

\autoref{fig:transform} presents an example. The first column shows two clean input images. The second column shows the injected backdoor samples that are transformed from clean inputs using the Toaster filter. Observe that the injected backdoor samples have dark orange color in the middle and lighter color for the surrounding areas. A straightforward design of the transformation layer is to use a traditional convolutional layer to transform the clean feature representation. The convolution operation denotes a uniform transformation, where all the values on a feature map is computed by a same kernel. However, this is undesirable for expressing the backdoor discussed above. The third column in~\autoref{fig:transform} denotes the generated samples by using a traditional convolutional layer. Observe that the color changes are uniform for different regions, failing to produce the orange color region in the middle. We hence propose to divide a feature map into a set of regions and apply different convolutional kernels on different regions. We call it {\em regional transformation}. The last column in~\autoref{fig:transform} presents the results of using regional transformation for generating backdoor samples. Observe that comparing to the images in the third row, the regional transformation is able to produce the orange color in the middle and lighter color in the surrounding areas.
Note that the Toaster filter is only one of the cases where
pervasive backdoors manipulate different regions of the input using different transformations. \ours{} does not follow the style of Toaster filter. We use the above case to demonstrate that \ours{} can model various pervasive backdoors using regional transformation. Please see another case study for the SIG attack~\cite{barni2019new} in Appendix~\ref{app:regional_case}.

\begin{figure}[t]
    \centering
    \includegraphics[width=0.9\columnwidth]{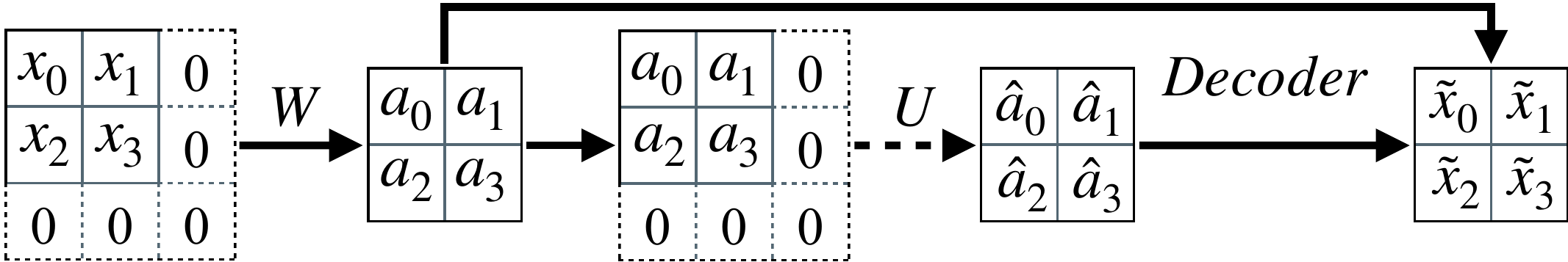}
    \caption{Example for transforming internal values}
    \label{fig:internal_conv}
\end{figure}

We formally define the regional transformation in the following. Assume the input feature representation  $a' \in \sR^{C' \times W' \times H'}$ (features before transformation), and a set of convolutional variables $\mU \in \sR^{z \times z \times C' \times C' \times m \times m}$, where $z \times z$ is the number of convolutional variables and $m$ is the convolutional kernel size. We first divide $a'$ into a set of regions with the size of $\frac{W'}{z} \times \frac{H'}{z}$, denoted as $w \times h$. The feature representation $a'$ can hence be reshaped to $a' \in \sR^{z \times z \times C' \times w \times h}$. The transformed feature representation $a''$ is obtained as follows.

\vspace*{-12pt}
{\small
\begin{equation}
\begin{split}
    a'' &= \Psi (a') = 
        \begin{bmatrix}
            r_{0,0} & r_{0,1} & ... & r_{0,z-1} \\
            r_{1,0} & r_{1,1} & ... & r_{1,z-1} \\
            ... & ... & ... & ... \\
            r_{z-1,0} & r_{z-1,1} & ... & r_{z-1,z-1} \\
        \end{bmatrix}, \\
    r_{i,j} &= \mU[i,j] \otimes a'[i, j],
\end{split}
\end{equation}}
where $r_{i,j}$ denotes the transformed region $(i, j)$ and $\otimes$ denotes the convolutional operation. Observe that each region $a'[i,j]$ is transformed by a convolutional variable $\mU[i,j]$. These regions will be placed in their original positions after the transformation. Note that the transformed feature representation $a''$ has the same number of channels as $a'$ such that it can be properly decoded by the decoder to the input space.

As a region of the feature representation is transformed by a convolutional variable $\mU$, we study the property of such a operation for expressing pervasive backdoor behaviors.
Assume a $2 \times 2$ input region $\mX$ on the left of \autoref{fig:internal_conv} and a convolution kernel parameterized by $\mW \in \R^{2 \times 2}$. Zero-padding is used (demonstrated by the dotted cells). Output values can be derived from the values in the region and the parameter values through the following equations.

\vspace*{-18pt}
{\small
\begin{align*}
a_0 &= w_0 \cdot x_0 + w_1 \cdot x_1 + w_2 \cdot x_2 + w_3 \cdot x_3 \\
a_1 &= w_0 \cdot x_1 + w_2 \cdot x_3 \\
a_2 &= w_0 \cdot x_2 + w_1 \cdot x_3 \\
a_3 &= w_0 \cdot x_3
\numberthis
\end{align*}}
\vspace*{-18pt}

\noindent
Here, we leave the activation functions out for discussion simplicity. Suppose the pervasive backdoor applies adversarial perturbation $\mathbf{\vdelta}$ on the region $\mX$. That is, $x'_i = x_i + \delta_i, i \in \{0, 1, 2, 3\}$. The feature representation for the backdoor sample region $\mA'$ is hence the following.

\vspace*{-18pt}
{\small
\begin{align*}
a'_0 &= w_0 \cdot (x_0 + \delta_0) + w_1 \cdot (x_1 + \delta_1) + w_2 \cdot (x_2 + \delta_2) + w_3 \cdot (x_3 + \delta_3) \\
a'_1 &= w_0 \cdot (x_1 + \delta_1) + w_2 \cdot (x_3 + \delta_3) \\
a'_2 &= w_0 \cdot (x_2 + \delta_2) + w_1 \cdot (x_3 + \delta_3) \\
a'_3 &= w_0 \cdot (x_3 + \delta_3)
\numberthis
\end{align*}}
\vspace*{-18pt}

Our goal is to derive backdoor samples from benign inputs. That is, we apply the convolutional operation on the benign feature representation to produce the backdoor representation. Here, we use a convolutional kernel $\mU \in \R^{2 \times 2}$ for analysis simplicity. Applying the kernel on the normal representation $\mA$ (see the middle part of \autoref{fig:internal_conv}) produces the following.

\vspace*{-18pt}
{\small
\begin{align*}
\hat{a}_0 &= u_0 \cdot a_0 + u_1 \cdot a_1 + u_2 \cdot a_2 + u_3 \cdot a_3 \\
          &= u_0 \cdot (w_0 x_0 + w_1 x_1 + w_2 x_2 + w_3 x_3) + u_1 \cdot (w_0 x_1 + w_2 x_3) \\
          &\quad + u_2 \cdot (w_0 x_2 + w_1 x_3) + u_3 \cdot w_0 x_3 \\
\hat{a}_1 &= u_0 \cdot a_1 + u_2 \cdot a_3
          = u_0 \cdot (w_0 x_1 + w_2 x_3) + u_2 \cdot w_0 x_3 \\
\hat{a}_2 &= u_0 \cdot a_2 + u_1 \cdot a_3
          = u_0 \cdot (w_0 x_2 + w_1 x_3) + u_1 \cdot w_0 x_3 \\
\hat{a}_3 &= u_0 \cdot a_3
          = u_0 \cdot w_0 x_3
\numberthis
\end{align*}}
Let $\mA' = \hat{\mA}$ and we have
{\small
\begin{align*}
\delta_0 &= (u_0 - 1) \cdot x_0 + u_1 \cdot x_1 + u_2 \cdot x_2 + u_3 \cdot x_3 \\
\delta_1 &= (u_0 - 1) \cdot x_1 + u_2 \cdot x_3 \\
\delta_2 &= (u_0 - 1) \cdot x_2 + u_1 \cdot x_3 \\
\delta_3 &= (u_0 - 1) \cdot x_3
\numberthis \label{eq:perturb_relation}
\end{align*}}
\vspace*{-18pt}

As observed in \autoref{fig:transform}, pervasive backdoors transform inputs based on each original pixel value and do not introduce abrupt value changes in the neighborhood of each pixel (within the region). That is, each pixel perturbation introduced by the backdoor transformation correlates to the original value of its corresponding pixel and the neighboring pixels. This can be expressed by our method as show in Equation~\ref{eq:perturb_relation}. For instance, the perturbation on the first pixel $\delta_0$ is a portion ($u_0 - 1$) of the corresponding pixel $x_0$ and also the linear combination of neighboring pixels ($u_1 x_1 + u_2 x_2 + u_3 x_3$). The scale of the perturbation is parameterized by our convolutional transformation $\mU$. It can be properly modeled during our backdoor generation using the gradient information from the subject model. The above analysis only considers one convolutional kernel in our transformation layer within the region for discussion simplicity. In practice, for example, the feature representation has 64 channels and each channel is associated with one kernel, which gives us 64 different combinations of neighboring pixels for each region.

\noindent
\textbf{Loss Terms.}
\autoref{fig:backdoor_gen} shows three loss terms. The SSIM score and the content loss are introduced to constrain the transformations on the inputs. In other words, it is desired to have generated backdoor samples retaining most main features and similar to the original inputs as pervasive backdoors preserve main contents (see~\autoref{fig:backdoor_examples}).

\vspace*{-16pt}
{\small
\begin{align}
    \gL_{SSIM} &= SSIM (x, \hat{x}) \\
    \gL_{content} &= MSE (a, \hat{a}) = \frac{1}{N} \sum_{i=0}^{N-1} (a_i - \hat{a}_i)^2
\end{align}}
\vspace*{-16pt}

The cross entropy loss $\gL_{CE} (y_t, \hat{y})$ is to induce the desired misclassification to the target label $y_t$. Other than the above three loss functions, we also use another two loss terms to improve the quality of generated backdoors as follows.

\vspace*{-18pt}
{\small
\begin{align}
    \gL_{norm} &= \frac{1}{C} \sum^{C} | \mu_b - \bar{\mu}_\mX | + \frac{1}{C} \sum^{C} | \sigma_b - \bar{\sigma}_\mX | \\
    \gL_{smooth} &= MSE \big(\hat{x}, AvgPool(\hat{x}) \big)
\end{align}}
\vspace*{-18pt}

\noindent
Loss term $\gL_{norm}$ is to reduce the difference between the backdoor statistics (i.e., mean and standard deviation) and the average statistics across all the samples $\mX$ (in the generation set) on each channel. This avoids the generated backdoor samples too far away from the distribution of input samples. Loss term $\gL_{smooth}$ smooths the local area of pixel changes, preventing abrupt pixel changes on the backdoor samples. Function $AvgPool$ is an average pooling operation, where each pixel value is replaced by the average of its neighboring pixels (e.g, in a $3 \times 3$ region).

Our final loss function for generating pervasive backdoors is in the following.

\vspace*{-18pt}
{\small
\begin{equation}
    \gL = \gL_{CE} + \alpha (\lambda_0 \gL_{content} + \lambda_1 \gL_{SSIM} + \lambda_2 \gL_{smooth} + \gL_{norm})
\end{equation}}
\vspace*{-18pt}

\noindent
We dynamically adjust the weight parameter $\alpha$ to balance the misclassification goal and the backdoor quality. We empirically set $\lambda_0 = 0.001$, $\lambda_1 = 100$, and $\lambda_2 = 0.05$ such that all the loss terms are at the same scale.

\section{Evaluation}
\label{sec:eval}

We evaluate \ours{} on three different scenarios including enlarging class distances against natural pervasive backdoors, eliminating injected backdoors in poisoned models, and reducing false positives for backdoor scanners. The experiments are conducted on six standard benchmarks, tens of pre-trained models from the TrojAI competition~\cite{TrojAI:online}, and six types of pervasive backdoors. We also carry out an adaptive attack to further test \ours{} and an ablation study to understand the effects of different design choices of \ours{}.

\subsection{Experiment Setup}
\label{sec:setup}

\noindent
\textbf{Datasets and Models.}
Six standard datasets and eight well-known model structures as well as a number of pre-trained models from the TrojAI competition are adopted for the evaluation. In the experiments of enlarging class distance, we employ CIFAR-10~\cite{krizhevsky2009learning}, SVHN~\cite{netzer2011reading}, LISA~\cite{mogelmose2012vision}, and GTSRB~\cite{stallkamp2012man}. Due to the page limit, the results on LISA and GTSRB are presented in Appendix~\ref{app:harden_distance}. Various model structures such as ResNet~\cite{he2016deep}, Network in Network (NiN)~\cite{lin2013network}, and VGG~\cite{simonyan2014very} are used. In addition, we evaluate on hardening adversarially trained models as well. For eliminating injected backdoors, we evaluate on six types of pervasive backdoors such as DFST~\cite{cheng2021deep}, Blend attack~\cite{chen2017targeted}, Sinusoidal Signal attck (SIG)~\cite{barni2019new}, WaNet~\cite{nguyen2021wanet}, Clean Label attack~\cite{turner2018clean}, and filter attack~\cite{liu2019abs}. More details can be found in Appendix~\ref{app:setup}.

\noindent
\textbf{Baselines.}
Three existing techniques discussed in Section~\ref{sec:motivation} are used as the baselines: adversarial training, UAP, and \moth{}. We also consider six state-of-the-art backdoor removal techniques, namely, Neural Cleanse (NC)~\cite{wang2019neural}, Neural Attention Distillation (NAD)~\cite{li2021neural}, Adversarial Neuron Pruning (ANP)~\cite{wu2021adversarial}, Artificial Brain Stimulation (ABS)~\cite{liu2019abs}, Fine-pruning~\cite{liu2018fine}, and Mode Connectivity Repair (MCR)~\cite{zhao2020bridging} as the additional baselines. Please see detailed description of these baselines in Appendix~\ref{app:baselines}.

\noindent
\textbf{Metrics.}
The following metrics are used in the evaluation. The normal functionalities are measured using the predication accuracy on the test set. The model robustness is measured using Projected Gradient Descent (PGD)~\cite{madry2018towards} within the given $\normmax$ bound (see details in Appendix~\ref{app:adv_setting}). The efficiency of different methods is measured by their training time in minutes. As discussed in Section~\ref{sec:motivation}, hardening class distances enhances the model's robustness against backdoor attacks. We hence use the relative improvement of pair-wise class distance as the metric. Specifically, for every class pair, we calculate the ratio of the distance difference between the hardened model and the original model over the original model's distance. The average of all class pairs is served as the final metric. The definition is as follows.

\vspace*{-15pt}
{\small
\begin{equation}
    \frac{1}{n \times (n - 1)} \sum_{i=1}^n \sum_{j=1, j \neq i}^n \frac{\hat{d}_{i \rightarrow j} - d_{i \rightarrow j}}{d_{i \rightarrow j}},
\end{equation}}
where $n$ is the number of classes; $d_{i \rightarrow j}$ and $\hat{d}_{i \rightarrow j}$ are the class distances from $i$ to $j$ for the original model and the hardened model, respectively. The metric is similar to that used in \moth{}. The difference is that here we use our pervasive backdoor for the measurement.
This is because just like in adversarial examples, $\normlzero$ norm cannot be used to measure $\normmax$ attacks (and vice versa). Analogously, patch backdoors and pervasive backdoors shall be studied and measured differently. Also, \ours{} has high ASRs as shown in~\autoref{fig:distance_asr} and hence is suitable to measure the distance.
The details of measuring the distance are in Appendix~\ref{app:metrics}. For poisoned models, we use the attack success rate (ASR) of the injected backdoor as the metric, which is the percentage of backdoor samples correctly classified to the target label.

\begin{table}[t]
    \caption{Evaluation on hardening class distance for naturally trained models. {\normalfont Columns 1-5 denote different datasets (D), models (M), hardening methods, model accuracy and robustness. Column 6 presents the runtime in minutes. Column 7 shows the average class distance across all class pairs. Column 8 denotes the improvement of pairwise class distance by different techniques compared to that of original models (Natural). Columns 9-10 (ADeg. and RDeg.) show the degradation of test accuracy and model robustness, respectively.}}
    \label{tab:harden_nat_main}
    \scriptsize
    \centering
    \tabcolsep=3.9pt
    \begin{tabular}{cc*{8}{r}}
        \toprule
        D & M & Method & Acc. & Rob. & Time & Dist. & Increase & ADeg. & RDeg. \\

        \midrule
        \multirow{20}{*}[-0.1in]{\rotatebox{90}{CIFAR-10}}
        & \multirow{7}{*}{\rotatebox{90}{ResNet20}}
        &   Natural & 91.52\% & 0.00\% & 56.77 & 22.62 & - & - & - \\

        & & NC      & 89.95\% & 0.00\%  & 84.96 & 24.11 & 6.01\% & 1.57\% & 0.00\% \\

        & & NAD     & 91.09\% & 0.00\%  &  3.06 & 22.57 & -0.29\% & 0.43\% & 0.00\% \\
        
        & & ANP     & 90.31\% & 0.00\%  &  0.95 & 22.69 & 0.29\% & 1.21\% & 0.00\% \\

        & & UAP     & 90.04\% & 0.00\%  & 243.11 & 23.98 & 6.64\% & 1.48\% & 0.00\% \\

        & & \moth{} & 90.34\% & 0.00\%  & 29.68 & 25.92 & 14.04\% & 1.18\% & 0.00\% \\
        
        & & \ours{} & 90.31\% & 0.00\%  & 16.73 & \textbf{34.60} & \textbf{52.66\%} & 1.21\% & 0.00\% \\

        \cmidrule{2-10}
        & \multirow{7}{*}{\rotatebox{90}{NiN}}
        &   Natural & 88.09\% & 0.20\% & 68.30 & 21.18 & - & - & - \\

        & & NC      & 87.18\% & 0.50\% & 40.38 & 21.46 & 1.55\% & 0.91\% & 0.00\% \\

        & & NAD     & 83.68\% & 0.30\% & 1.14 & 21.52 & 2.32\% & 4.41\% & 0.00\% \\
        
        & & ANP     & 85.88\% & 0.20\% & 0.41 & 20.96 & -0.98\% & 2.21\% & 0.00\% \\

        & & UAP     & 86.61\% & 0.30\% & 196.67 & 21.13 & 0.03\% & 1.48\% & 0.00\% \\

        & & \moth{} & 86.81\% & 0.00\% & 36.63 & 23.21 & 10.04\% & 1.28\% & 0.20\% \\
        
        & & \ours{} & 86.35\% & 0.20\% & 21.02 & \textbf{25.95} & \textbf{22.86\%} & 1.74\% & 0.00\% \\

        \cmidrule{2-10}
        & \multirow{7}{*}{\rotatebox{90}{VGG19}}
        &   Natural & 92.30\% & 0.20\% & 68.42 & 27.60 & - & - & - \\

        & & NC      & 91.23\% & 0.90\% & 134.08 & 27.19 & -0.87\% & 1.07\% & 0.00\% \\

        & & NAD     & 91.51\% & 0.20\% & 3.83 & 28.58 & 3.40\% & 0.79\% & 0.00\% \\
        
        & & ANP     & 90.48\% & 0.20\% & 1.04 & 28.57 & 4.90\% & 1.82\% & 0.00\% \\

        & & UAP     & 90.78\% & 0.90\% & 226.55 & 25.35 & -5.25\% & 1.52\% & 0.00\% \\

        & & \moth{} & 91.48\% & 0.20\% & 44.80 & 26.64 & -2.09\% & 0.82\% & 0.00\% \\
        
        & & \ours{} & 90.88\% & 1.50\% & 19.00 & \textbf{37.88} & \textbf{39.72\%} & 1.42\% & 0.00\% \\

        \midrule
        \multirow{14}{*}[-0.03in]{\rotatebox{90}{SVHN}}
        & \multirow{7}{*}{\rotatebox{90}{NiN}}
        &   Natural & 95.61\% & 6.50\% & 10.50 & 23.43 & - & - & - \\

        & & NC      & 94.39\% & 6.10\% & 24.75 & 23.33 & 0.53\% & 1.22\% & 0.40\% \\

        & & NAD     & 92.48\% & 9.80\% & 1.42 & 25.03 & 7.54\% & 3.13\% & 0.00\% \\
        
        & & ANP     & 93.72\% & 4.10\% & 0.55 & 23.14 & -0.50\% & 1.89\% & 2.40\% \\

        & & UAP     & 94.63\% & 4.30\% & 45.47 & 22.54 & -3.18\% & 0.98\% & 2.20\% \\

        & & \moth{} & 94.99\% & 6.60\% & 47.77 & 23.37 & 0.54\% & 0.62\% & 0.00\% \\
        
        & & \ours{} & 94.08\% & 7.10\% & 18.33 & \textbf{37.55} & \textbf{61.47\%} & 1.53\% & 0.00\% \\

        \cmidrule{2-10}
        & \multirow{7}{*}{\rotatebox{90}{ResNet32}}
        &   Natural & 95.15\% & 0.20\% & 26.70 & 36.40 & - & - & - \\

        & & NC      & 94.09\% & 0.30\% & 31.59 & 33.87 & -6.13\% & 1.06\% & 0.00\% \\

        & & NAD     & 93.91\% & 0.40\% & 0.89 & 34.36 & -5.12\% & 1.24\% & 0.00\% \\
        
        & & ANP     & 93.99\% & 0.20\% & 3.61 & 37.59 & 4.19\% & 1.16\% & 0.00\% \\

        & & UAP     & 93.16\% & 0.10\% & 228.95 & 39.51 & 9.85\% & 1.99\% & 0.10\% \\

        & & \moth{} & 94.49\% & 7.90\% & 172.63 & 27.48 & -23.90\% & 0.66\% & 0.00\% \\
        
        & & \ours{} & 93.84\% & 0.70\% & 29.15 & \textbf{87.20} & \textbf{142.73\%} & 1.31\% & 0.00\% \\

        \midrule
        \multicolumn{2}{c}{\multirow{7}{*}{Average}}
        &   Natural & 92.53\% & 1.42\% & 46.14 & 26.24 & - & - & - \\

        & & NC      & 91.37\% & 1.56\% & 63.15 & 25.99 & 0.22\% & 1.16\% & 0.00\% \\

        & & NAD     & 90.53\% & 2.14\% & 2.07 & 26.41 & 1.57\% & 2.00\% & 0.00\% \\
        
        & & ANP     & 90.88\% & 0.94\% & 1.31 & 26.59 & 1.58\% & 1.66\% & 0.48\% \\

        & & UAP     & 91.04\% & 1.14\% & 188.15 & 26.50 & 1.62\% & 1.49\% & 0.28\% \\

        & & \moth{} & 91.62\% & 2.94\% & 66.30 & 25.32 & -0.27\% & 0.91\% & 0.00\% \\
        
        & & \ours{} & 91.09\% & 1.90\% & 21.86 & \textbf{44.64} & \textbf{63.89\%} & 1.44\% & 0.00\% \\

        \bottomrule
    \end{tabular}
    \vspace{-5pt}
\end{table}

\subsection{Evaluation on Enlarging Class Distance}
\label{sec:eval_harden}

Three settings are considered in studying the performance of various techniques on enlarging class distances, including naturally trained models, adversarially trained models, and pre-trained models from TrojAI round 4. 
The results are shown in~\autoref{tab:harden_nat_main},~\autoref{tab:harden_adv_main}, and~\autoref{tab:harden_trojai} (in Appendix). Due to the page limit, the results on LISA and GTSRB are presented in~\autoref{tab:harden_nat_app} and~\autoref{tab:harden_adv_app} in Appendix~\ref{app:harden_distance}. 
As hardening using UAP requires training a model from scratch~\cite{shafahi2020universal}, it is only compared with other techniques on naturally trained models.

\noindent\textbf{Results on Naturally Trained Models.}
From~\autoref{tab:harden_nat_main}, observe that \ours{} can enlarge the class distance from 26.24 to 44.64 on average (63.89\% improvement), with less than 1.5\% accuracy degradation and no robustness loss. \ours{} has the largest distance improvement on ResNet32 with SVHN, with 142.73\% enlargement. The three backdoor removal techniques, NC, NAD, and ANP, have limited performance on the class distance, with less than 2\% improvement on average. In some cases, they even decrease the distance, for instance, -6.13\% and -5.12\% improvements for NC and NAD, respectively, on hardening ResNet32 with SVHN. The largest distance improvements for NC, NAD, and ANP are 6.01\%, 7.54\%, 4.9\%, respectively, lower than the smallest improvement of 22.86\% by \ours{}. The above observation is reasonable as these backdoor removal techniques were originally designed specifically for eliminating injected backdoors in poisoned models. They ought to have little impact on benign models and hence cannot enlarge distances for normal class pairs. NC takes 63.15 minutes on average to harden a model, even longer than training the original model (46.14 minutes). NAD and ANP have low time costs, with 2.07 and 1.31 minutes overhead respectively. UAP can increase the class distance for 9.85\% on ResNet32 with SVHN. But its performance on other datasets and models are still limited, with an average of 1.62\% enlargement. As discussed in Section~\ref{sec:motivation}, UAP is bounded by $\normlp$ and hard to derive feasible perturbations for flipping the majority of samples from a victim class to the target class. Increasing the $\normlp$ bound will cause large accuracy degradation, which is not desired for preserving normal functionalities. In addition, UAP is extremely slow, taking 188.15 minutes on average, which is 4x longer than the natural training.

\begin{table}[t]
    \caption{Evaluation on hardening class distance for adversarially trained models}
    \label{tab:harden_adv_main}
    \scriptsize
    \centering
    \tabcolsep=2.8pt
    \begin{tabular}{cc*{8}{r}}
        \toprule
        D & M & Method & Acc. & Rob. & Time & Dist. & Increase & ADeg. & RDeg. \\

        \midrule
        \multirow{18}{*}[-0.05in]{\rotatebox{90}{CIFAR-10}}
        & \multirow{6}{*}{\rotatebox{90}{ResNet20}}
        &   Adversarial & 74.87\% & 42.80\% & 462.50 & 32.28 & - & - & - \\

        & & NC          & 74.96\% & 42.50\% & 78.73 & 32.92	& 2.48\% & 0.00\% & 0.30\% \\

        & & NAD         & 74.29\% & 42.50\% & 2.61 & 32.56 & 1.08\% & 0.58\% & 0.30\% \\

        & & ANP         & 73.75\% & 40.90\% & 0.97 & 31.39 & -2.41\% & 1.12\% & 1.90\% \\

        & & \moth{}     & 74.04\% & 43.00\% & 38.71 & 33.72 & 4.76\% & 0.83\% & 0.00\% \\
        
        & & \ours{}     & 74.55\% & 42.00\% & 15.52 & \textbf{39.18} & \textbf{22.56\%} & 0.32\% & 0.80\% \\

        \cmidrule{2-10}
        & \multirow{6}{*}{\rotatebox{90}{NiN}}
        &   Adversarial & 73.94\% & 41.40\% & 430.39 & 30.28 & - & - & - \\

        & & NC          & 73.75\% & 38.00\% & 43.41 & 30.03 & -0.40\% & 0.19\% & 3.40\% \\

        & & NAD         & 73.61\% & 39.50\% & 1.26 & 30.12 & -0.27\% & 0.33\% & 1.90\% \\
        
        & & ANP         & 73.10\% & 36.00\% & 0.42 & 29.59 & -2.23\% & 0.84\% & 5.40\% \\

        & & \moth{}     & 73.47\% & 41.10\% & 25.57 & 32.36 & 7.27\% & 0.47\% & 0.30\% \\
        
        & & \ours{}     & 73.01\% & 40.60\% & 14.11 & \textbf{41.29} & \textbf{37.81\%} & 0.93\% & 0.80\% \\

        \cmidrule{2-10}
        & \multirow{6}{*}{\rotatebox{90}{VGG19}}
        &   Adversarial & 76.00\% & 41.30\% & 1166.18 & 34.50 & - & - & - \\

        & & NC          & 75.37\% & 41.60\% & 141.38 & 34.31 & -0.49\% & 0.63\% & 0.00\% \\

        & & NAD         & 73.81\% & 39.90\% & 4.12 & 37.45 & 9.35\% & 2.19\% & 1.40\% \\
        
        & & ANP         & 75.69\% & 38.50\% & 1.07 & 32.95 & -4.11\% & 0.31\% & 2.80\% \\

        & & \moth{}     & 75.08\% & 42.50\% & 81.54 & 39.45 & 13.99\% & 0.92\% & 0.00\% \\
        
        & & \ours{}     & 75.68\% & 42.30\% & 18.21 & \textbf{43.31} & \textbf{25.63\%} & 0.32\% & 0.00\% \\

        \midrule
        \multirow{12}{*}[-0.03in]{\rotatebox{90}{SVHN}}
        & \multirow{6}{*}{\rotatebox{90}{NiN}}
        &   Adversarial & 91.63\% & 51.20\% & 132.00 & 45.99 & - & - & - \\

        & & NC          & 92.02\% & 41.50\% & 27.38 & 31.66 & -27.12\% & 0.00\% & 9.70\% \\

        & & NAD         & 88.73\% & 31.10\% & 1.42 & 42.92 & -0.96\% & 2.90\% & 20.10\% \\
        
        & & ANP         & 91.45\% & 46.90\% & 0.56 & 36.13 & -19.34\% & 0.18\% & 4.30\% \\

        & & \moth{}     & 91.40\% & 52.30\% & 51.42 & 38.40 & -14.43\% & 0.24\% & 0.00\% \\
        
        & & \ours{}     & 91.45\% & 50.20\% & 17.17 & \textbf{67.60} & \textbf{48.12\%} & 0.18\% & 1.00\% \\

        \cmidrule{2-10}
        & \multirow{6}{*}{\rotatebox{90}{ResNet32}}
        &   Adversarial & 92.94\% & 58.30\% & 300.08 & 47.54 & - & - & - \\

        & & NC          & 92.25\% & 49.90\% & 43.20 & 33.18 & -20.29\% & 0.69\% & 8.40\% \\

        & & NAD         & 90.02\% & 48.70\% & 5.81 & 48.25 & 15.85\% & 2.92\% & 9.60\% \\
        
        & & ANP         & 91.70\% & 51.50\% & 2.33 & 35.91 & -16.34\% & 1.24\% & 6.80\% \\

        & & \moth{}     & 92.52\% & 58.90\% & 111.74 & 35.44 & -17.10\% & 0.42\% & 0.00\% \\
        
        & & \ours{}     & 93.18\% & 58.10\% & 32.80 & \textbf{56.54} & \textbf{32.57\%} & 0.00\% & 0.20\% \\

        \midrule
        \multicolumn{2}{c}{\multirow{6}{*}{Average}}
        &   Adversarial & 81.88\% & 47.00\% & 498.23 & 38.12 & - & - & - \\

        & & NC          & 81.67\% & 42.70\% & 66.82 & 32.42 & -9.16\% & 0.21\% & 4.30\% \\

        & & NAD         & 80.09\% & 40.34\% & 3.04 & 38.26 & 5.01\% & 1.79\% & 6.66\% \\
        
        & & ANP         & 81.14\% & 42.76\% & 1.07 & 33.19 & -8.89\% & 0.74\% & 4.24\% \\

        & & \moth{}     & 81.30\% & 47.56\% & 61.80 & 35.87 & -1.10\% & 0.58\% & 0.00\% \\

        & & \ours{}     & 81.57\% & 46.64\% & 19.56 & \textbf{49.58} & \textbf{33.34\%} & 0.30\% & 0.36\% \\

        \bottomrule
    \end{tabular}
    \vspace{-5pt}
\end{table}

\moth{} is a state-of-the-art hardening technique designed for backdoor defense. It can improve the class distance by more than 10\% in two cases (ResNet20 and NiN on CIFAR-10). However, \moth{} performs poorly on ResNet32 with SVHN, with -23.90\% enlargement. Note that the distance measured in this paper is different from that in \moth{}. We use the difference of feature representations from the encoder between backdoor samples and their clean counterparts, whereas \moth{} leverages the trigger size in the input space (similar to the number of perturbed pixels), which is not suitable for measuring pervasive backdoors as they tend to alter all input pixels. The low performance of \moth{} on SVHN might be due to the characteristics of the task. SVHN is a dataset for recognizing digital numbers from $0$ to $9$. The hardening procedure of \moth{} relies on the backdoors generated by NC~\cite{wang2019neural}, which tends to produce backdoors with a small number of pixels for models on SVHN. Pervasive backdoors aim to transform the whole input instead of modifying a few pixels, which makes hardening techniques such as \moth{} targeting pixel-level backdoors insufficient. The time cost of \moth{} is 66.30 minutes, similar to training a natural model. By and large, \ours{} has the largest class distance improvement for the evaluated models, substantially outperforming baselines. \ours{} has a reasonable hardening cost, with less than half of the natural training (21.86 minutes on average). 
In Appendix~\ref{app:neuron_collapse}, we also demonstrate that leveraging the phenomenon of Neural Collapse~\cite{papyan2020prevalence} dose not improve the distance.

\noindent\textbf{Results on Adversarially Trained Models.}
Similar observations can be made on adversarially trained models in~\autoref{tab:harden_adv_main}. The clean accuracy of adversarailly trained models is usually lower than that of naturally trained ones. Although the class distance of adversarially trained models is much larger, they are still not optimal. \ours{} is able to further enlarge the distance by 33.34\% on average, with only 0.30\% and 0.36\% degradation on accuracy and robustness, respectively. NC and ANP have worse performance on adversarially trained models than on natural ones, with -9.16\% and -8.89\% enlargement, respectively. In addition, they both have more than 4\% robustness degradation, which is critical to adversarially trained models. NAD has larger improvement on two cases: VGG19 with CIFAR-10 (9.35\%) and ResNet32 with SVHN (15.85\%). The performance gain, however, is at the expense of much larger accuracy and robustness degradations (2.19\%-2.92\%) and (1.40\%-9.60\%), respectively, which is undesirable. \moth{} has a reasonable performance on CIFAR-10, improving the class distance by from 4.76\% to 13.99\%. The results on SVHN are inferior, which is consistent with the observation on naturally trained models and discussed earlier. The hardening time costs of all the methods are similar to those on naturally trained models.

\begin{figure}
    \centering
    \includegraphics[width=0.92\columnwidth]{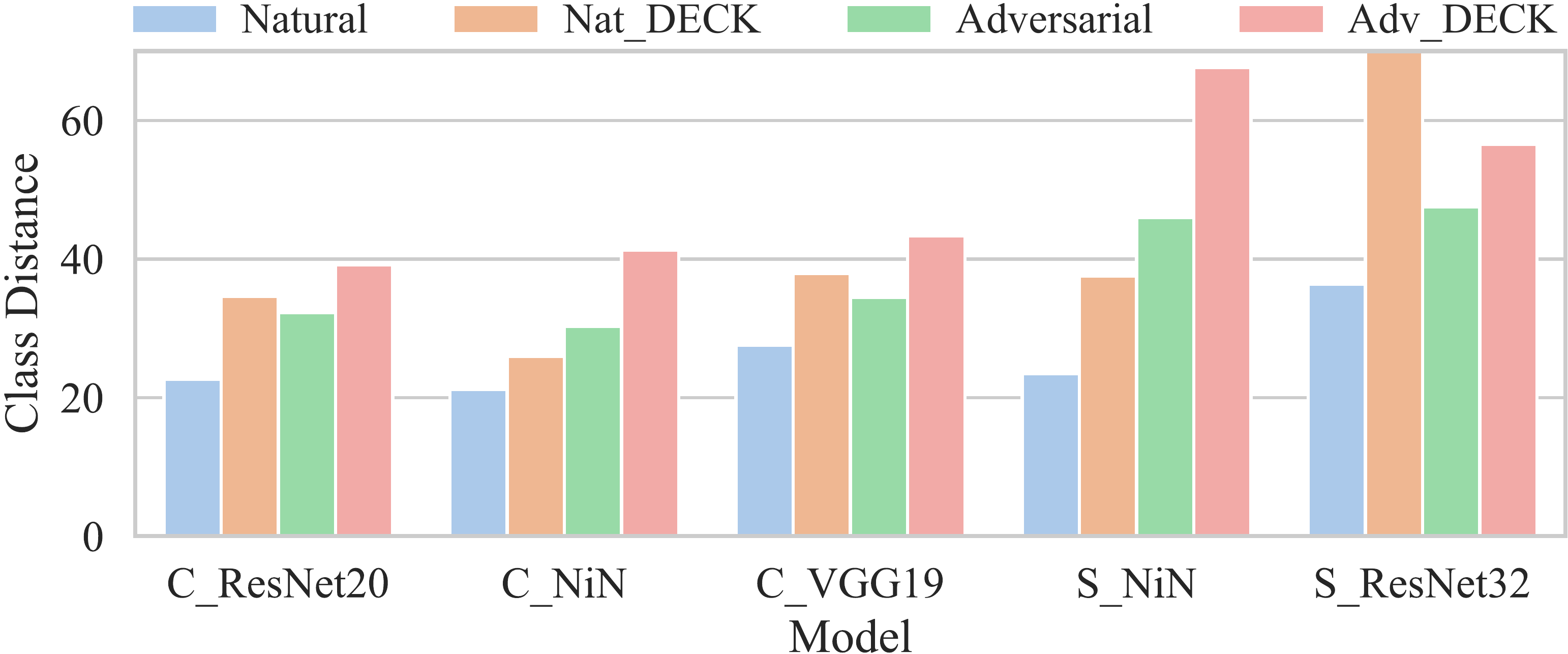}
    \caption{Comparison of the average class distance between naturally trained and adversarially trained models. {\normalfont The first letter of the model name denotes the dataset (C for CIFAR-10 and S for SVHN).}}
    \label{fig:dist_comparison}
\end{figure}

\noindent\textbf{Comparison of Naturally and Adversarially Trained Models.}
We further study the class distances of naturally trained and adversarially trained models as well as models hardened by \ours{}. The average class distances of different models are shown in~\autoref{fig:dist_comparison}. The x-axis denotes the dataset and model structure, and the y-axis denotes the class distance. Observe that all the adversarially trained models (green bars) have larger distances than their natural counterparts (blue bars). \ours{} can improve the class distance for both types of models. Particularly, three naturally trained models hardened by \ours{} (C\_ResNet20, C\_VGG19, and S\_ResNet32) have even larger class distances than adversarially trained models. The accuracy of \ours{}-hardened versions of these models (91.68\%) is much higher than adversarially trained models (81.27\%) on average. This renders adversarial training is insufficient of achieving the optimal class distance improvement even with the sacrifice of clean accuracy. \ours{} can obtain much larger distances starting from the adversarially trained models as shown by the red bars in~\autoref{fig:dist_comparison} with negligible degradation on accuracy (0.30\%) and robustness (0.36\%). In other words, \ours{} can provide a much better extra protection against pervasive backdoor attacks with very limited sacrifice.

\noindent\textbf{Results on Pre-trained Models.}
The evaluation on pre-trained models from TrojAI round 4 also exhibits similar findings. The results are shown in~\autoref{tab:harden_trojai} in Appendix due to the page limit. The two backdoor removal techniques NAD and ANP have limited performance, with 7.45\% and 3.66\% improvement on the class distance on average, respectively. \moth{} obtains a reasonable result with 34.42\% distance enlargement on average. Their improvements however are still inferior to \ours{} that can enlarge the distance by 88.81\%, from 4.01 to 6.93. Note that the pre-trained TrojAI models have already been hardened by adversarial training, indicating the effectiveness of \ours{} in hardening class distances for various settings.

\begin{table}[t]
    \caption{Comparison of distance measures by different backdoor generation methods. The fifth and seventh columns show the average class distance measured by \ours{} and filter backdoor, respectively.}
    \label{tab:distance_filter}
    \scriptsize
    \centering
    \tabcolsep=4.5pt
    \begin{tabular}{cc*{6}{r}}
        \toprule
        D & M & Method & Accuracy & Dist.$_\text{\ours{}}$ & Increase$_\text{\ours{}}$ & Dist.$_\text{filter}$ & Increase$_\text{filter}$ \\

        \midrule
        \multirow{7}{*}{\rotatebox{90}{CIFAR-10}}
        & \multirow{7}{*}{\rotatebox{90}{ResNet20}}
        & Natural       & 91.52\% & 22.62 &      - & 39.04 & - \\

        & & NC          & 89.95\% & 24.11 &  6.01\% & 43.46 & 9.67\% \\

        & & NAD         & 91.09\% & 22.57 & -0.29\% & 39.55 & -6.63\% \\
        
        & & ANP         & 90.31\% & 22.69 &  0.29\% & 33.45 & -7.02\% \\
        
        & & UAP         & 90.04\% & 23.98 &  6.64\% & 41.92 & 6.87\% \\
        
        & & \moth{}     & 90.34\% & 25.92 & 14.04\% & 42.78 & 9.58\% \\

        & & \ours{}     & 90.31\% & \textbf{34.60} & \textbf{52.66\%} & \textbf{58.14} & \textbf{20.42\%} \\

        \midrule
        \multirow{7}{*}{\rotatebox{90}{SVHN}}
        & \multirow{7}{*}{\rotatebox{90}{ResNet32}}
        &   Natural & 95.15\% & 36.40 &        - & 23.30 & - \\

        & & NC      & 94.09\% & 33.87 &  -6.13\% & 20.42 & -8.94\% \\

        & & NAD     & 93.91\% & 34.36 &  -5.12\% & 15.97 & -35.36\% \\
        
        & & ANP     & 93.99\% & 37.59 &   4.19\% & 26.38 & 10.59\% \\

        & & UAP     & 93.16\% & 39.51 &   9.85\% & 34.93 & 25.79\% \\

        & & \moth{} & 94.49\% & 27.48 & -23.90\% & 11.20 & -69.47\% \\
        
        & & \ours{} & 93.84\% & \textbf{87.20} & \textbf{142.73\%} & \textbf{41.72} & \textbf{148.35\%} \\
        
        \bottomrule
    \end{tabular}
    \vspace{-5pt}
\end{table}

\noindent\textbf{Study of Distance Measure.}
The class distance in the above evaluation is measured by \ours{} as it can achieve high ASRs as shown in~\autoref{fig:distance_asr}. To demonstrate the stability of the measure, we also study using another backdoor generation method, i.e., filter backdoor for measuring the distance. Note that most existing pervasive backdoor attacks require data poisoning and hence cannot be adapted to measure the distance for a given pre-trained model. Two naturally trained models are employed for the study and the results are demonstrated in~\autoref{tab:distance_filter}. Columns 4-6 are copied from~\autoref{tab:harden_nat_main}, where columns 5-6 show the distances measured by \ours{} and the relative improvements over the original model. Columns 7-8 display the distance results measured by filter backdoor. Observe that the distances measured by both \ours{} and filter backdoor have similar results. Baselines such as NC, \moth{} can only enlarge the distance for one of the two evaluated models. NAD fails to improve the distance for both models. UAP has relatively better performance with an average of 8.25\% and 16.33\% improvement measured by \ours{} and filter backdoor, respectively. Models hardened by \ours{} have the largest distance improvement (97.70\% and 84.39\% on average measured by our backdoor and filter backdoor, respectively). The distance measure by \ours{} is general and can characterize model's robustness against pervasive backdoor attacks. We also study the correlation of distance measures by \moth{} and \ours{} in Appendix~\ref{app:distance_correlation}, demonstrating that they focus on different aspects of class distance.

\subsection{Evaluation on Eliminating Injected Pervasive Backdoors}
\label{sec:eval_removal}

One of our most notable design goals is to use the transformation layer and the decoder to model a large spectrum of (existing) pervasive attacks. As such, hardening using our backdoor can improve resilience to all these attacks.
We investigate six types of pervasive backdoors and apply \ours{} as well as the state-of-the-art backdoor removal techniques to study whether injected backdoors can be eliminated. We follow the same setup as in existing works~\cite{li2021neural,tao2022model} by using only 5\% of the original training set to eliminate backdoors. Seven existing backdoor removal techniques, namely, Standard Finetuning, NAD~\cite{li2021neural}, ANP~\cite{wu2021adversarial}, ABS~\cite{liu2019abs}, Fine-pruning~\cite{liu2018fine}, MCR~\cite{zhao2020bridging}, and \moth{}~\cite{tao2022model}, are considered as the baselines. Standard Finetuning is a standard method originally proposed for transfer learning. It updates a pre-trained model's weights with a small learning rate on the training set. We leverage the finetuning baseline setting in NAD~\cite{li2021neural}, which adopts data augmentation techniques including random crop, horizontal flipping, and Cutout~\cite{devries2017improved} during training.
ABS is able to reverse engineer feature space attack, such as simple filter backdoors. We hence leverage ABS to first invert a filter-like trigger pattern and apply it on clean samples. The poisoned model is then finetuned on these trigger-stamped inputs to unlearn injected backdoors.
The descriptions of NAD, ANP, ABS, Fine-pruning, MCR, and \moth{} can be found in Section~\ref{sec:motivation} and Section~\ref{sec:setup}. We follow the original papers and the official repositories of the baselines, and carefully tune their settings to obtain the best results.
Due to the page limit, the results of Fine-pruning and MCR are shown in~\autoref{tab:backdoors_app} and discussed in Appendix~\ref{app:removal}.

\begin{table*}[t]
    \caption{Evaluation on backdoor attacks}
    \label{tab:backdoors}
    \scriptsize
    \centering
    \tabcolsep=1.8pt
    \begin{tabular}{cll*{14}{r}}
        \toprule
        \multirow{2}{*}[-0.03in]{\parbox{0.5in}{\centering Backdoor Attack}} & \multirow{2}{*}[-0.03in]{Dataset} & \multirow{2}{*}[-0.03in]{Model} & \multicolumn{2}{c}{Original} & \multicolumn{2}{c}{Standard Finetune} & \multicolumn{2}{c}{NAD} & \multicolumn{2}{c}{ANP} & \multicolumn{2}{c}{ABS} & \multicolumn{2}{c}{\moth{}} & \multicolumn{2}{c}{\ours{}} \\

        \cmidrule(lr){4-5} \cmidrule(lr){6-7} \cmidrule(lr){8-9} \cmidrule(lr){10-11} \cmidrule(lr){12-13} \cmidrule(lr){14-15} \cmidrule(lr){16-17}
        & & & Accuracy & ASR & Accuracy & ASR & Accuracy & ASR & Accuracy & ASR & Accuracy & ASR & Accuracy & ASR & Accuracy & ASR \\

        \midrule
        \multirow{8}{*}{DFST}
        & \multirow{4}{*}{CIFAR-10}
        & ResNet32 Detox1
        & 89.95\% & 97.60\% & 87.30\% & 70.00\% & 88.00\% & 60.44\% & 83.34\% & 20.67\% & 86.97\% & 84.11\% & 88.90\% & 65.44\% & 88.22\% & \underline{14.22\%} \\
        
        & & ResNet32 Detox3
        & 90.93\% & 95.33\% & 89.20\% & 63.89\% & 85.78\% & 17.33\% & 83.68\% & 90.11\% & 90.74\% & 37.78\% & 89.53\% & 54.44\% & 88.11\% & \underline{12.22\%} \\
        
        & & VGG13 Detox1
        & 90.34\% & 95.89\% & 86.07\% & 93.56\% & 87.24\% & 24.56\% & 86.92\% & 89.56\% & 88.71\% & 33.78\% & 87.26\% & 5.33\% & 88.03\% & \underline{2.00\%} \\
        
        & & VGG13 Detox3
        & 91.29\% & 97.44\% & 89.55\% & 66.11\% & 87.09\% & 55.33\% & 88.46\% & 96.11\% & 88.67\% & 66.67\% & 89.91\% & 51.22\% & 89.08\% & \underline{5.67\%} \\
        
        \cmidrule{2-17}
        
        & \multirow{4}{*}{STL-10}
        & ResNet32 Detox1
        & 75.74\% & 97.67\% & 70.64\% & 70.64\% & 70.92\% & 44.00\% & 65.71\% & 90.11\% & 72.26\% & 68.56\% & 71.97\% & 60.89\% & 72.10\% & \underline{2.67\%} \\
        
        & & ResNet32 Detox3
        & 76.45\% & 99.00\% & 71.30\% & 93.22\% & 72.21\% & 89.11\% & 71.89\% & 69.56\% & 71.95\% & 97.78\% & 72.09\% & 71.56\% & 72.86\% & \underline{4.78\%} \\
        
        & & VGG13 Detox1
        & 72.18\% & 98.67\% & 70.11\% & 84.78\% & 68.91\% & 86.67\% & 68.88\% & 98.11\% & 68.46\% & 48.44\% & 69.86\% & 62.67\% & 68.61\% & \underline{5.89\%} \\
        
        & & VGG13 Detox3
        & 72.09\% & 98.89\% & 70.42\% & 97.33\% & 68.91\% & 81.00\% & 65.70\% & 97.33\% & 67.29\% & 37.56\% & 67.54\% & 86.56\% & 69.89\% & \underline{12.33\%} \\
        
        \midrule
        \multirow{2}{*}{Blend}
        & CIFAR-10 & ResNet20
        & 90.96\% & 99.96\% & 90.33\% & 84.92\% & 86.81\% & 3.94\% & 85.20\% & 6.22\% & 89.41\% & 5.66\% & 85.44\% & 12.26\% & 89.08\% & \underline{0.00\%} \\
        
        & SVHN & NiN
        & 94.10\% & 92.37\% & 92.70\% & 0.54\% & 94.40\% & 0.33\% & 92.67\% & 0.41\% & 91.76\% & 10.66\% & 94.41\% & \underline{0.20\%} & 94.56\% & 0.85\% \\
        
        \midrule
        \multirow{2}{*}{SIG}
        & CIFAR-10 & ResNet20
        & 83.38\% & 93.30\% & 88.65\% & 59.40\% & 85.84\% & 9.44\% & 80.02\% & 37.44\% & 86.82\% & 85.29\% & 80.39\% & 17.72\% & 86.91\% & \underline{3.97\%} \\
        
        & SVHN & NiN
        & 95.48\% & 92.46\% & 95.11\% & 41.60\% & 94.21\% & 0.68\% & 90.10\% & 20.91\% & 93.47\% & 55.02\% & 94.90\% & 0.69\% & 93.96\% & \underline{0.46\%} \\

        \midrule
        \multirow{3}{*}{WaNet}
        & CIFAR-10 & ResNet18
        & 94.15\% & 99.55\% & 93.58\% & 80.71\% & 91.37\% & 0.87\% & 91.38\% & \underline{0.11\%} & 89.61\% & 1.88\% & 92.65\% & 0.62\% & 91.12\% & 0.64\% \\
        
        & GTSRB & ResNet18
        & 99.01\% & 98.94\% & 96.80\% & 48.75\% & 94.96\% & 0.02\% & 97.38\% & 0.00\% & 98.51\% & \underline{0.00\%} & 97.72\% & 0.01\% & 97.70\% & 0.30\% \\
        
        & CelebA & ResNet18
        & 78.99\% & 99.08\% & 78.89\% & 21.35\% & 76.57\% & 15.34\% & 76.79\% & 14.22\% & 75.56\% & 21.69\% & 77.80\% & 8.91\% & 77.57\% & \underline{8.12\%} \\
        
        \midrule
        Clean Label & CIFAR-10 & ResNet18
        & 87.60\% & 98.36\% & 85.74\% & 97.34\% & 84.44\% & 7.47\% & 86.18\% & 4.46\% & 85.33\% & 7.18\% & 83.70\% & 8.34\% & 85.67\% & \underline{4.22\%} \\
        
        \midrule
        
        \multicolumn{3}{c}{\textbf{Average}}
        & \textbf{86.42\%} & \textbf{97.16\%} & \textbf{84.77\%} & \textbf{67.13\%} & \textbf{83.60\%} & \textbf{31.03\%} & \textbf{82.14\%} & \textbf{45.96\%} & \textbf{84.10\%} & \textbf{41.38\%} & \textbf{84.00\%} & \textbf{31.68\%} & \textbf{84.59\%} & \textbf{\underline{4.90\%}} \\

        \bottomrule
    \end{tabular}
    \vspace{-15pt}
\end{table*}

The results on five pervasive backdoors are presented in~\autoref{tab:backdoors} and on the filter backdoor in~\autoref{fig:poison_asr} and~\autoref{fig:poison_acc} (in Appendix). In~\autoref{tab:backdoors}, columns 1-3 denote the attacks, datasets, and model structures. Columns 4-5 show the clean accuracy of poisoned models and the attack success rate (ASR) of injected backdoors. The following columns present the accuracy and ASR of models after applying the different techniques. Deep Feature Space Trojan attack (DFST)~\cite{cheng2021deep} introduces a detoxification procedure by iteratively training on ABS~\cite{liu2019abs} reverse-engineered backdoors to reduce the number of compromised neurons that can be leveraged by existing scanners for successful detection. We follow the original paper and evaluate on two settings with one round (Detox1) and three rounds (Detox3) of detoxification (the original paper~\cite{cheng2021deep} used at most three rounds). Observe that on CIFAR-10, Standard Finetuning, NAD, ABS, and \moth{} can only reduce the ASR of DFST on ResNet32 Detox1 from 97.60\% to more than 60\%. ANP is able to reduce the ASR to 20.67\% but with the cost of large accuracy degradation from 89.95\% to 83.34\%. \ours{}, on the other hand, can reduce the ASR to 14.22\% with only 1.73\% accuracy degradation. NAD, ABS, and \moth{}{} have better performance on VGG13 Detox1, reducing the ASR from 95.89\% to 24.56\%, 33.78\%, and 5.33\%, respectively. However, with the increase of detoxification rounds, they can only reduce the ASR to more than 50\%. ANP has poor performance on VGG13 with less than 10\% ASR reduction for both one round and three rounds of detoxifications. \ours{} can consistently eliminate DFST backdoors, e.g., less than 15\% ASR on ResNet32 and 6\% on VGG13. 
The evaluation on the STL-10 dataset shows similar results. The baselines have limited performances with the best result of reducing the ASR to 37.56\% on VGG13 Detox3 by ABS.
\ours{}, on the other hand, can consistently remove injected backdoors to achieve as low as 2.67\% ASR. In the worst case (on VGG13 Detox3), \ours{} can still reduce the ASR to 12.33\%.

Blend attack uses random small perturbation patterns as the backdoor, which can be easily eliminated by all the evaluated techniques, except for Standard Finetuning on CIFAR-10. \ours{} can reduce the ASRs to less than 1\% for the two studied cases. We have similar observations for the results on SIG, WaNet, and Clean Label backdoors. Existing techniques can reduce the ASRs to a reasonable range (less than 20\% for most cases). \ours{} further lower the ASRs to less than 5\% for 5 out of 6 cases and 9\% for the remaining case (WaNet on CelebA), outperforming the state-of-the-art approaches. The last row of~\autoref{tab:backdoors} presents the average results on the five pervasive backdoors. Standard Finetuning can only reduce the ASR from 97.16\% to 67.13\% with 1.65\% accuracy degradation. NAD and \moth{} can eliminate backdoors to having around 30\% ASR with around 2.5\% accuracy degradation. ANP is able to reduce the ASR to 45.96\% with 4.28\% accuracy degradation.
ABS has a similar defense performance by reducing the ASR to 41.38\% with 2.32\% accuracy drop.
Fine-pruning and MCR can eliminate backdoors to having around 50\% ASR with around 4\% accuracy degradation. Due to the page limit, the breakdown of their defense results is elaborated in Appendix~\ref{app:removal}.
\ours{} has the largest ASR reduction from 97.16\% to 4.90\% and the accuracy degradation is only 1.83\%.

Similar observations can be made for the evaluation on the filter backdoor, which is shown in~\autoref{fig:poison_asr}. The x-axis and y-axis denote the model IDs and the ASR, respectively. The bars in the light/dark colors show the ASR of injected backdoors before/after applying backdoor-erasing techniques. Observe that Standard finetuning (blue bars) can only repair half of the evaluated models (17 out of the 34 models). This is expected as backdoor attacks include clean data together with backdoored samples during training. Finetuning only on clean samples may not eliminate the backdoor patterns that have already been learned by poisoned models. NAD leverages the teacher-student structure and treats the model from Standard Finetuning as the teacher network. Its performance hence is limited by Standard Finetuning. This can be observed from the orange bars in~\autoref{fig:poison_asr}. NAD is only able to eliminate five more backdoors (with a total of 21 models). ANP has limited performance on TrojAI models, with only 15 poisoned models being repaired. The TrojAI poisoned models were trained by NIST~\cite{TrojAI:online} and different training strategies including random transformations, adversarial training, etc., are employed to make injected backdoors more robust and hard to  detect. These strategies may reduce the sensitivity of individual neurons on backdoor patterns. ANP is hence not able to identify compromised neurons and fails to remove injected backdoors. This observation is consistent with the results on DFST backdoors that apply detoxification to reduce compromised neurons. 
ABS can only repair 15 models. As the injected backdoors in TrojAI models are label-specific, ABS may not be able to identify the correct victim-target class pair. The inverted triggers fail to expose the injected backdoor behaviors. Unlearning on those triggers hence cannot repair models.

\begin{figure*}[t]
    \centering
    \includegraphics[width=\textwidth]{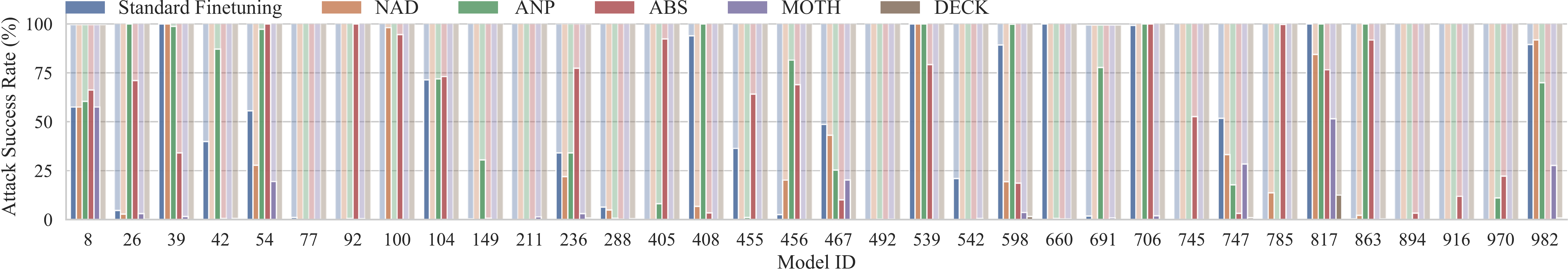}
    \caption{Attack success rate of poisoned models before (light color) and after (dark color) repair}
    \label{fig:poison_asr}
    \vspace{-10pt}
\end{figure*}

\moth{} can eliminate more backdoors than other baselines with 28 fixed models. As discussed in Section~\ref{sec:motivation}, pervasive backdoors perturb all pixels on the input and are dynamic, which is different from \moth{}'s threat model that focuses on patch-like static backdoors. It can raise the bar for pervasive backdoor to some extent but still fails to repair 6 TrojAI poisoned models. \ours{}, on the other hand, can eliminate all the backdoors with an average ASR down to 0.55\%, outperforming the others. The accuracies of poisoned models before and after repair are shown in~\autoref{fig:poison_acc} in Appendix. Overall, all the approaches incur very small accuracy degradation on average ($<0.3\%$), except for ANP (1.16\%).

\subsection{Reducing False Positives in Backdoor Scanning}

Backdoor scanning aims to distinguish trojaned models from benign ones. The TrojAI competition~\cite{TrojAI:online} aims to evaluate the performance of various backdoor scanners submitted by the performers. However, due to the existence of natural pervasive backdoors in benign models, the performers encounter the problem of false positives, which are benign models misclassified as trojaned ones. To alleviate the problem, the TrojAI organizer adopted adversarial training for constructing benign models in round 4. The top performer however still reports 14 false positives. In Section~\ref{sec:eval_harden}, we have demonstrated that \ours{} can effectively enlarge class distance of models for defending natural pervasive backdoors. Hence, an idea is to apply \ours{} on the 14 TrojAI (benign) models to check if they are still misclassified by scanners.
After \ours{}'s hardening, the accuracy degradation of these models is minimal (0.10\% on average). The distance improvement is similar to that evaluated in Section~\ref{sec:eval_harden} and hence omitted. We then run the top performer's model scanner (downloaded from the official repository~\cite{scanner}) on the round 4 dataset (with the 14 hardened models). There is no false positive reported by the scanner and no degradation on the true positives. The result demonstrates the capability of \ours{} in assisting backdoor scanning.
This echoes our argument in Section~\ref{sec:motivation} that in the future, published pre-trained models should undergo hardening similar to what we are proposing here. As such, any intentionally compromised models can be easily detected.

\subsection{Adaptive Attack}
\label{sec:adaptive}

\begin{table}[t]
    \centering
    \caption{Adaptive Attack}
    \begin{tabular}{crrrr}
        \toprule
        \multirow{2}{*}[-0.03in]{Poisoning Rate} & \multicolumn{2}{c}{Backdoored} & \multicolumn{2}{c}{\ours{}} \\

        \cmidrule(lr){2-3} \cmidrule(lr){4-5}
        & Accuracy & ASR & Accuracy & ASR \\
        
        \midrule
        20\% & 83.64\% & 60.43\% & 83.94\% & 9.94\% \\
        
        50\% & 67.86\% & 74.94\% & 75.71\% & 4.42\% \\
        
        \bottomrule
    \end{tabular}
    \label{tab:adaptive_attack}
    \vspace{-5pt}
\end{table}

We conduct an adaptive attack by optimizing a trigger pattern while applying \ours{}, the adaptive knowledge, during pervasive poisoning. We use the cross entropy loss to ensure the pattern can induce the target label and the $\normltwo$ loss to minimize the perturbation magnitude. The experiment is conducted on a ResNet20 model with CIFAR-10. The results are shown in~\autoref{tab:adaptive_attack}. The first column shows the poisoning rate. The following columns show the accuracy and ASR for the backdoored models before and after repaired by \ours{}. Observe that using a 20\% poisoning rate, the backdoored model has 83.64\% accuracy and 60.43\% ASR. By increasing the poisoning rate to 50\%, the ASR is improved to 74.94\% with much more accuracy degradation (67.86\% accuracy). As \ours{} aims to increase class distance while the poisoning incurs a small distance to the target class, the two contradicting goals make the accuracy much lower than a clean model (91.52\%) and the ASR low as well. By applying \ours{} to the poisoned models, the ASR drops to 9.94\%/4.42\% without accuracy degradation (83.94\%/75.71\%) as shown in the last two columns in~\autoref{tab:adaptive_attack}, delineating \ours{}'s resilience to adaptive attacks.

\subsection{Ablation Study}
\label{sec:ablation}

\ours{} features a few important design choices.
We study different decoder designs as well as the position of the feature representation in the encoder. We also investigate the effects of four loss terms used in pervasive backdoor generation. We observe that the best class distance enlargement of existing decoders is still 16.36\% lower than ours. Using a deeper layer of the encoder for feature extraction does not boost the distance improvement. The four loss terms are all important, with $\gL_{SSIM}$ and $\gL_{norm}$ the crucial ones. Please see detailed discussion in Appendix~\ref{app:ablation}.
\section{Related Work}
\label{sec:related}

Backdoor attacks have been a critical security threat to deep learning models. In early works, a static trigger pattern is commonly used for stamping on training samples to poison the model, such as patch attacks~\cite{GuLDG19,chen2017targeted}. Recently, pervasive backdoors have been explored by researchers. Different from patch attacks, pervasive backdoors utilizes a backdoor function that transforms the whole input with input-specific perturbation. They are visually similar to their benign counterparts, making them stealthy and hard to detect. We have studied and evaluated six representative pervasive backdoor attacks in this paper, including DFST~\cite{cheng2021deep}, Blend attack~\cite{chen2017targeted}, SIG attack~\cite{barni2019new}, WaNet~\cite{nguyen2021wanet}, Clean Label attack~\cite{turner2018clean}, and Filter attack~\cite{liu2019abs}.

Defense techniques against backdoor attacks can be categorized into backdoor input detection, certified robustness, backdoor scanning, and backdoor removal. Backdoor input detection aims to detect inputs stamped with backdoor triggers, such as STRIP~\cite{gao2019strip}, Spectral Signatures~\cite{tran2018spectral}, and so on~\cite{ma2019nic,chen2018detecting,li2020rethinking,liu2017neural,chan2019poison,du2019robust,veldanda2020nnoculation}. Certified robustness provides certification to the classification results of individual samples, asserting the results can be trusted even in the presence of backdoors~\cite{mccoyd2020minority,xiang2021patchguard,xiang2021patchcleanser,jia2020certified}. Backdoor scanning focuses on identify whether a given model has been injected with backdoor. They leverage trigger inversion techniques to reverse-engineer injected backdoors~\cite{wang2019neural,liu2019abs,shen2021backdoor}, train a two-class detection network~\cite{kolouri2020universal,xu2019detecting}, utilize adversarial perturbation to observe model behaviors~\cite{huang2020one,zhang2020cassandra,wang2020practical}, and so on~\cite{tang2021demon}. These techniques are orthogonal to \ours{} as they do not consider natural backdoors. Backdoor removal aims to eliminate injected backdoors in poisoned models~\cite{liu2018fine,zeng2020deepsweep}. State-of-the-art techniques leverage knowledge distillation to remove backdoors~\cite{li2021neural}, prune out potentially compromised neurons~\cite{wu2021adversarial}, or harden models by enlarging input-space class distances~\cite{tao2022model}. Our evaluation in Section~\ref{sec:eval_removal} demonstrates the effectiveness of \ours{} in eliminating backdoors, surpassing the state-of-the-arts. Researchers also study the backdoor learning curves to assist backdoor defense~\cite{cina2021backdoor}.
\section{Conclusion}
We propose a general pervasive backdoor attack that models a spectrum of pervasive backdoors, which can be employed to harden models. The evaluation on various datasets and model structures demonstrates that our technique can improve model robustness against pervasive backdoors and reduce the attack success rate of six pervasive backdoor attacks from 99.06\% to 1.94\%, outperforming existing state-of-the-art hardening techniques.



\bibliographystyle{plain}
\bibliography{sections/references}

\clearpage
\newpage

\section*{Appendix}
\setcounter{section}{0}
\renewcommand*{\theHsection}{appendix.\the\value{section}}

\section{Essence of the Various Upsampling Methods}
\label{app:upsampling}

\begin{figure}
    \centering
    \includegraphics[width=0.99\columnwidth]{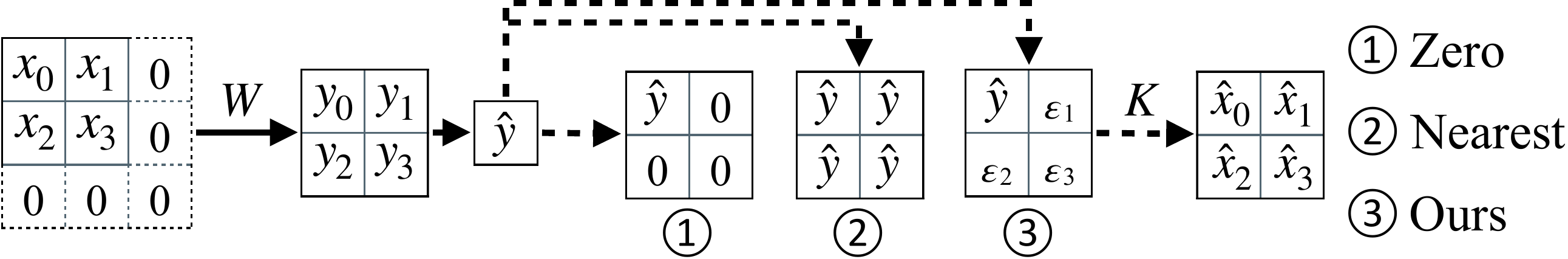}
    \caption{Example for different upsampling methods}
    \label{fig:decoder_up_math}
\end{figure}

Assume a $2 \times 2$ input matrix $\mathbf{X}$ on the left of \autoref{fig:decoder_up_math} and a convolution kernel parameterized by $\mathbf{W} \in \mathbb{R}^{2 \times 2}$. Zero-padding is used (demonstrated by the dotted cells). Output values can be derived from the values in the matrix and the parameter values through the following equations on the top (lines 1-4). The inverse equations are shown on the bottom (lines 5-8).

\vspace*{-10pt}
{\small
\begin{align*}
y_0 &= w_0 \cdot x_0 + w_1 \cdot x_1 + w_2 \cdot x_2 + w_3 \cdot x_3 \\
y_1 &= w_0 \cdot x_1 + \cancel{w_1 \cdot 0} \; + w_2 \cdot x_3 + \cancel{w_3 \cdot 0} \\
y_2 &= w_0 \cdot x_2 + w_1 \cdot x_3 + \cancel{w_2 \cdot 0} \; + \cancel{w_3 \cdot 0} \\
y_3 &= w_0 \cdot x_3 + \cancel{w_1 \cdot 0} \; + \cancel{w_2 \cdot 0} \; + \cancel{w_3 \cdot 0} \\
x_0 &= \tfrac{1}{w_0} y_0 - \tfrac{w_1}{w_0^2} y_1 - \tfrac{w_2}{w_0^2} y_2 - \tfrac{w_0 w_3 - 2 w_1 w_2}{w_0^3} y_3 \\
x_1 &= \tfrac{1}{w_0} y_1 - \tfrac{w_2}{w_0^2} y_3 \\
x_2 &= \tfrac{1}{w_0} y_2 - \tfrac{w_1}{w_0^2} y_3 \\
x_3 &= \tfrac{1}{w_0} y_3
\numberthis \label{eq:downsample}
\end{align*}}

Here, we leave the activation functions out for discussion simplicity. The zero terms are crossed out. Downsampling is normally employed after convolutional operations to reduce data dimensions. We use max-pooling with a stride of 2 as an illustrative example. The resulted value (after downsampling) is hence $\hat{y} = \max (y_i), i \in \{0, 1, 2, 3\}$. The goal of upsampling is to recover values $x_i$ from $\hat{y}$. Firstly, the upsampling methods (\circled{1}-\circled{3} in \autoref{fig:decoder_up_math}) expand $\hat{y}$ to the original matrix shape in different ways. A transposed convolutional operation then follows right after that to recover the input values, i.e., $\hat{x}_0, ..., \hat{x}_3$. Suppose the transposed convolution kernel is parameterized by $\mathbf{K} \in \mathbb{R}^{2 \times 2}$. The upsampled values $\hat{x}_0, ..., \hat{x}_3$ can be formalized as follows for each method.

\vspace*{-10pt}
{\small
\begin{align*}
\hat{x}_0 = \circled{1} \; &k_0 \cdot \hat{y}
    & \circled{2} \; & k_0 \cdot \hat{y} \\
\hat{x}_1 = \quad \;\; &k_1 \cdot \hat{y} + \cancel{k_0 \cdot 0}
    & & (k_1 + k_0) \cdot \hat{y} \\
\hat{x}_2 = \quad \;\; &k_2 \cdot \hat{y} + \cancel{k_0 \cdot 0}
    & & (k_2 + k_0) \cdot \hat{y} \\
\hat{x}_3 = \quad \;\; &k_3 \cdot \hat{y} + \cancel{k_2 \cdot 0} + \cancel{...}
    & & (k_3 + k_2 + k_1 + k_0) \cdot \hat{y}
\end{align*}}
{\small
\begin{align*}
    \circled{3} \; & k_0 \cdot \hat{y} \\
    \quad \;\; & k_1 \cdot \hat{y} + k_0 \cdot \epsilon_1 \\
    \quad \;\; & k_2 \cdot \hat{y} + k_0 \cdot \epsilon_2 \\
    \quad \;\; & k_3 \cdot \hat{y} + k_2 \cdot \epsilon_1 + k_1 \cdot \epsilon_2 + k_0 \cdot \epsilon_3 
\numberthis \label{eq:upsample}
\end{align*}}

The constant $\epsilon_1$, $\epsilon_2$, and $\epsilon_3$ in \circled{3} are values randomly sampled from a Gaussian distribution $\mathcal{N} (0, \sigma^2)$, where $\sigma$ is trained from the activation values from the corresponding layer in the encoder. We use a zero-mean Gaussian as it is simpler and does not require training for the mean. The results of using trained mean are similar to using a zero-mean Gaussian and hence omitted. Observe that both upsampling methods \circled{1} and \circled{2} induce a fixed linear correlation among the decoded values $\hat{x}_i$ (through $\hat{y}$). For instance in method \circled{1}, $\frac{\hat{x}_2}{\hat{x}_1}\equiv \frac{k_2}{k_1}$. However, as we can see from Equation~\ref{eq:downsample}, such correlations do not exist in downsampling.

In our method \circled{3}, we mitigate such fixed correlations through randomization. Specifically, we leverage a Gaussian distribution to help approximating the possible disregarded values. In Equation~\ref{eq:upsample}, decoded values $\hat{x}_1, \hat{x}_2, \hat{x}_3$ have correlations with both the internal value $\hat{y}$ and the randomly sampled values $\epsilon_i \sim \mathcal{N} (0, \sigma^2)$ ($\sigma$ is collected from the activation values in the encoder). The magnitude of the randomization is controlled by the transposed convolution kernel $k_i$, allowing to express more complex relations. In addition, just like other randomization procedures (e.g., random dropout) used in deep neural network training, the randomization (in upsampling) also enhances the generality and robustness of the decoding kernels. During decoder training, we sample $\epsilon_i$ from Gaussian distributions for each upsampling layer. 

\section{Regional Transformation for SIG Attack}
\label{app:regional_case}

We use the SIG attack (the 4th column in~\autoref{fig:backdoor_examples}) that adds a strip-like pattern on images, as an example to study the capability of \ours{} in modeling various pervasive backdoors. As ABS~\cite{liu2019abs} can invert feature map based triggers, we employ it as a baseline to compare with \ours{}. Specifically, we apply ABS and \ours{} to invert triggers for a poisoned ResNet20 model on CIFAR-10. The ASR of the inverted trigger is 39.00\% by ABS and 97.00\% by \ours{}. We also calculate the MSE between original poisoned images and inverted poisoned images. The distances are 1480.46 for ABS, and 1073.41 for \ours{}, delineating \ours{} can better simulate pervasive backdoors.

\section{Details of Experiment Setup}
\label{app:setup}

\subsection{Datasets and Models}

\noindent
\textbf{CIFAR-10}~\cite{krizhevsky2009learning} is an object recognition dataset with 10 classification classes. It consists of 60,000 images and is divided into a training set (48,000 images), a validation set (2,000 images), and a test set (10,000 images).

\smallskip
\noindent
\textbf{SVHN}~\cite{netzer2011reading} is a dataset contains house digital numbers extracted from Google Street View images. It has 73,257 training images and 26,032 test images. We divide the original training set into 67,257 images for training and 6,000 images for validation.

\smallskip
\noindent
\textbf{STL-10}~\cite{coates2011analysis} is an image recognition dataset with 10 classification classes. It consists of 5,000 training images and 8,000 test images.

\smallskip
\noindent
\textbf{LISA}~\cite{mogelmose2012vision} is a U.S. traffic sign dataset with 47 different road signs. However, the number of images in different classes is not well balanced. Some classes have very few images, which is hard to train a reasonable model. We follow an existing work~\cite{eykholt2018robust} by choosing 18 most common classes based on their numbers of training images. We use 5,635 images for training, 704 images for validation, and 704 images for testing.

\smallskip
\noindent
\textbf{GTSRB}~\cite{stallkamp2012man} is a German traffic sign recognition dataset with 43 classes. We split the dataset into a training set (35,289 images), a validation set (3,920 images), and a test set (12,630 images).

\begin{figure*}[t]
    \centering
    \includegraphics[width=\textwidth]{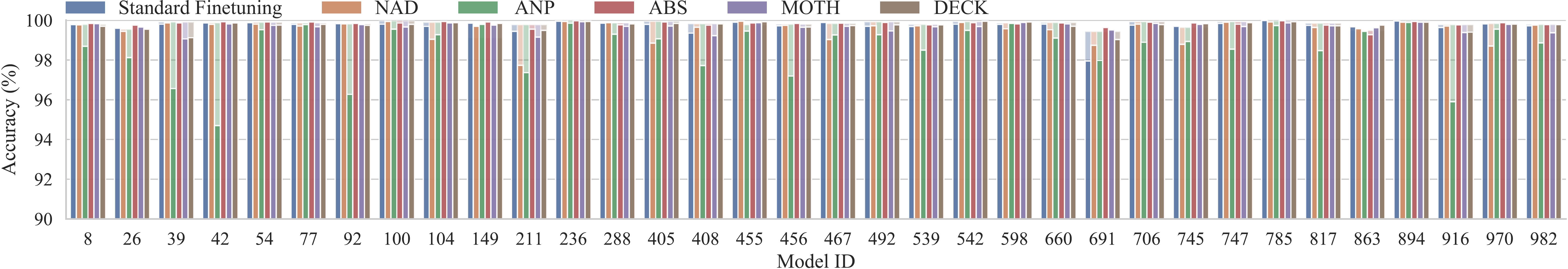}
    \caption{Accuracy of poisoned models before (light color) and after (dark color) repair}
    \label{fig:poison_acc}
\end{figure*}

\smallskip
\noindent
\textbf{CelebA}~\cite{liu2015deep} is a face attributes dataset. It contains 10,177 identities with 202,599 face images. Each image has an annotation of 40 binary attributes. We follow~\cite{nguyen2021wanet} to select 3 out of 40 attributes, i.e., Heavy Makeup, Mouth Slightly Open, and Smiling, and create an 8-class classification task.

\smallskip
\noindent
\textbf{TrojAI}~\cite{TrojAI:online} round 4 includes 16 types of model structures such as InceptionV3~\cite{szegedy2016rethinking}, DenseNet121~\cite{huang2017densely}, SqueezeNet~\cite{iandola2016squeezenet}, etc. The task of these models is to recognize synthetic street traffic signs with between 15 and 45 classes. Input images are constructed by compositing a foreground object, e.g., a synthetic traffic sign, with a random background images from five different dataset such as Cityscapes~\cite{cordts2016cityscapes}, KITTI~\cite{Geiger2013IJRR}, Swedish Roads~\cite{larsson2011correlating}, etc. A set of random transformations are applied during model training, such as blurring, lighting, shifting, titling, etc. Adversarial training such as PGD~\cite{madry2018towards} and FBF~\cite{wong2019fast} is also utilized to improve model quality. We use random seed $1030792629$ to select 10 pre-trained models from the list of benign models from TrojAI round 4~\cite{TrojAI:online}. For the filter attack, we use random seed $186270393$ to select 34 poisoned models.

\subsection{Baselines}
\label{app:baselines}
\noindent\textbf{NC}~\cite{wang2019neural} generates backdoors for each class and checks whether there exists an exceptionally small backdoor. If this is the case, it then injects the generated backdoor on 20\% of the available training set (10\% of the original training set) to retrain the subject model.

\noindent\textbf{NAD}~\cite{li2021neural} leverages the teacher-student structure to eliminate backdoors. It first finetunes the poisoned model on 5\% of the training set. It uses this finetuned model as the teacher network, and the poisoned model as the student network. It then aims to reduce the internal feature differences between the teacher network and the student network by updating the student network. Finally, NAD outputs the student network as the cleaned model.

\noindent\textbf{ANP}~\cite{wu2021adversarial} is based on the observation that backdoor related neurons are more sensitive to adversarial perturbations on their weights. It hence applies a mask on all the neurons in the model, adversarially perturbs neuron weights to increase the classification loss for a set of clean samples, and minimizes the size of mask. ANP then prunes neurons with small mask values, meaning that they have been compromised by backdoor attacks.

\noindent
\textbf{ABS}~\cite{liu2019abs} introduces a neuron stimulation analysis to expose abnormal behaviors of neurons in a deep neural network by increasing their activation values. Those neurons are regarded as compromised neurons and leveraged to reverse engineer backdoor triggers. ABS proposes a one-layer transformation to approximate/invert filter triggers. The inverted trigger is hence utilized to remove the injected backdoor in poisoned models following the unlearning procedure in NC~\cite{wang2019neural} (see the above description).

\noindent
\textbf{Fine-pruning}~\cite{liu2018fine} prunes neurons that have low activation values for a set of clean samples. It then finetune the pruned model on a small set of clean samples.

\noindent
\textbf{MCR}~\cite{zhao2020bridging} linearly interpolates the weight parameters of two models. It also includes a set of trainable parameters during the interpolation. Specifically, the following equation is used to build a new model $\phi_\theta(t)$.

\vspace*{-15pt}
{\small
\begin{equation}
    \phi_\theta (t) = (1 - t)^2 \omega_1 + 2t(1-t) \theta + t^2 \omega_2, \;\; 0 \leq t \leq 1,
\end{equation}}
where $t$ is the interpolation hyper-parameter ranging from $0$ to $1$. $\omega_1$ and $\omega_2$ are the weight parameters of two pre-trained models, which are fixed. $\theta$ is a set of trainable parameters that have the same shape of $\omega_1$ and $\omega_2$. For eliminating backdoors in poisoned models, MCR uses the poisoned model and its finetuned version as the two endpoints ($\omega_1$ and $\omega_2$) and trains $\theta$ on a small set of clean samples. The best $t$ is chosen for the interpolation based on the clean accuracy.

\subsection{Adversarial Training and Robustness}
\label{app:adv_setting}

PGD~\cite{madry2018towards} is leveraged for creating adversarially trained models in Section~\ref{sec:eval_harden}. We use $\normmax$ bound of $8/255$ for CIFAR-10, 0.03 for SVHN and GTSRB, and 0.1 for LISA. 
We follow the standard definition of $\normmax$ robustness by running PGD for multiple steps to attack the subject model~\cite{madry2018towards} within the above given bound. The robustness measures how many generated adversarial examples are correctly classified as the ground truth by the subject model.
For hardening with UAP, the $\normmax$ bound is determined based on the clean accuracy degradation. We use $\normmax$ bound of $4/255$ for CIFAR-10, 0.05 for SVHN, and 0.03 for LISA and GTSRB.

\begin{table}[t]
    \caption{Comparison of different methods on hardening class distance for naturally trained models (cont.)}
    \label{tab:harden_nat_app}
    \scriptsize
    \centering
    \tabcolsep=3.5pt
    \begin{tabular}{cc*{8}{r}}
        \toprule
        D & M & Method & Acc. & Rob. & Time & Dist. & Increase & ADeg. & RDeg. \\

        \midrule
        \multirow{7}{*}{\rotatebox{90}{CIFAR-10}}
        & \multirow{7}{*}{\rotatebox{90}{ResNet50}}
        &   Natural     & 92.71\% & 0.00\% & 74.11 & 23.05 & - & - & - \\
        
        & & NC          & 91.54\% & 0.00\% & 183.00 & 24.40 & 5.78\% & 1.17\% & 0.00\% \\

        & & NAD         & 92.65\% & 0.00\% & 6.01 & 23.10 & 0.48\% & 0.06\% & 0.00\% \\
        
        & & ANP         & 92.27\% & 0.00\% & 4.66 & 23.35 & 1.35\% & 0.44\% & 0.00\% \\

        & & UAP         & 91.83\% & 0.70\% & 345.61& 24.10 & 5.08\% & 0.88\% & 0.00\% \\

        & & \moth{}     & 91.25\% & 0.00\% & 51.69 & 24.84 & 8.22\% & 1.46\% & 0.00\% \\
        
        & & \ours{}     & 91.43\% & 0.00\% & 21.49 & \textbf{42.42} & \textbf{83.96\%} & 1.28\% & 0.00\% \\

        \midrule
        \multirow{14}{*}[-0.03in]{\rotatebox{90}{LISA}}
        & \multirow{7}{*}{\rotatebox{90}{CNN}}
        &   Natural     & 97.30\% & 1.00\% & 0.15 & 42.51 & - & - & - \\

        & & NC          & 82.67\% & 1.14\% & 8.27 & 46.86 & 15.46\% & 14.63\% & 0.00\% \\

        & & NAD         & 95.45\% & 0.86\% & 0.15 & 44.79 & 8.58\% & 1.85\% & 0.14\% \\
        
        & & ANP         & 96.45\% & 1.14\% & 0.07 & 45.68 & 11.30\% & 0.85\% & 0.00\% \\

        & & UAP         & 95.60\% & 3.57\% & 1.79 & 39.35 & -2.40\% & 1.70\% & 0.00\% \\

        & & \moth{}     & 96.31\% & 1.29\% & 13.09 & 50.16 & 18.69\% & 0.99\% & 0.00\% \\
        
        & & \ours{}     & 96.73\% & 0.71\% & 3.82 & \textbf{82.56} & \textbf{111.38\%} & 0.57\% & 0.29\% \\

        \cmidrule{2-10}
        & \multirow{7}{*}{\rotatebox{90}{ResNet20}}
        &   Natural     & 98.86\% & 0.00\% & 1.70 & 29.79 & - & - \\

        & & NC          & 98.44\% & 0.00\% & 34.42 & 29.49 & 0.81\% & 0.42\% & 0.00\% \\

        & & NAD         & 96.59\% & 0.00\% & 0.72 & 30.99 & 5.51\% & 2.27\% & 0.00\% \\
        
        & & ANP         & 96.88\% & 0.00\% & 1.16 & 28.74 & -1.47\% & 1.98\% & 0.00\% \\

        & & UAP         & 96.16\% & 0.00\% & 6.33 & 29.58 & 2.67\% & 2.70\% & 0.00\% \\

        & & \moth{}     & 98.30\% & 0.14\% & 38.75 & 27.02 & -5.55\% & 0.57\% & 0.00\% \\
        
        & & \ours{}     & 98.44\% & 0.00\% & 4.42 & \textbf{34.71} & \textbf{17.65\%} & 0.42\% & 0.00\% \\

        \midrule
        \multirow{7}{*}{\rotatebox{90}{GTSRB}}
        & \multirow{7}{*}{\rotatebox{90}{NiN}}
        &   Natural     & 95.28\% & 13.90\% & 4.60 & 29.28 & - & - & - \\

        & & NC          & 95.65\% & 13.30\% & 110.51 & 29.28 & 0.15\% & 0.00\% & 0.60\% \\

        & & NAD         & 93.63\% & 12.20\% & 0.51 & 32.74 & 14.96\% & 1.65\% & 1.70\% \\
        
        & & ANP         & 94.05\% & 13.30\% & 0.43 & 29.13 & 0.68\% & 1.23\% & 0.60\% \\

        & & UAP         & 95.25\% & 15.80\% & 20.55 & 30.47 & 7.44\% & 0.03\% & 0.00\% \\

        & & \moth{}     & 95.55\% & 17.40\% & 74.01 & 30.75 & 8.00\% & 0.00\% & 0.00\% \\
        
        & & \ours{}     & 96.33\% & 12.80\% & 45.71 & \textbf{33.33} & \textbf{16.48\%} & 0.00\% & 1.10\% \\

        \midrule
        \multicolumn{2}{c}{\multirow{7}{*}{Average}}
        &   Natural     & 96.04\% & 3.73\% & 20.14 & 31.16 & - & - & - \\

        & & NC          & 92.08\% & 3.61\% & 84.05 & 32.51 & 5.55\% & 3.96\% & 0.12\% \\

        & & NAD         & 94.58\% & 3.26\% & 1.85 & 32.91 & 7.38\% & 1.46\% & 0.46\% \\
        
        & & ANP         & 94.91\% & 3.61\% & 1.58 & 31.73 & 2.97\% & 1.13\% & 0.11\% \\

        & & UAP         & 94.71\% & 5.02\% & 93.57 & 30.88 & 3.20\% & 1.33\% & 0.00\% \\

        & & \moth{}     & 95.35\% & 4.71\% & 44.39 & 33.19 & 7.34\% & 0.69\% & 0.00\% \\
        
        & & \ours{}     & 95.73\% & 3.38\% & 18.86 & \textbf{48.26} & \textbf{57.37\%} & 0.31\% & 0.35\% \\

        \bottomrule
    \end{tabular}
\end{table}

\subsection{Pervasive Backdoor Attacks}

\noindent
\textbf{Deep Feature Space Trojan (DFST)}~\cite{cheng2021deep} leverages a generative adversarial network (GAN) to inject a certain style (e.g., sunrise color style) to given training samples. It also introduces a detoxification procedure by iteratively training on ABS~\cite{liu2019abs} reverse-engineered backdoors to reduce the number of compromised neurons that can be leveraged by existing scanners for successful detection. We follow the original paper and poison models with two settings: one-round detoxification and three-rounds detoxification.

\smallskip
\noindent
\textbf{Blend attack}~\cite{chen2017targeted} injects a random perturbation pattern on the training samples of non-target classes and changes the ground truth labels of these samples to the target class (label 0). We use the random pattern reported in the original paper and poison 10\% of the non-target-class training data with a blend ratio of $\alpha = 0.2$.

\smallskip
\noindent
\textbf{Sinusoidal Signal attck (SIG)}~\cite{barni2019new} injects a strip-like pattern on the training samples of the target class and retains the original ground truth labels. We follow the setting in the original paper and generate the backdoor pattern using the horizontal sinusoidal function with $\Delta = 20$ and $f = 6$. We use label 0 as the target class and poison 8\% of the training data in the target class.

\smallskip
\noindent
\textbf{WaNet}~\cite{nguyen2021wanet} uses elastic image warping that deforms an image by applying the distortion transformation (e.g., distorting straight lines) as the backdoor. We download three backdoored models from the official repository~\cite{nguyen2021wanet}, which are trained on CIFAR-10, GTSRB, and CelebA, respectively.

\smallskip
\noindent
\textbf{Clean Label attack}~\cite{turner2018clean} generates adversarial perturbations on the training samples in the target class using an adversarailly trained model. It then injects a $2 \times 2$ grid at the top left corner of the target-class inputs and retain their ground truth labels. We use $\normmax$ bound of $8/255$ for crafting adversarial perturbations, use label 3 as the target class, and poison 50\% of the training data in the target class following the official repository~\cite{cleanlabel}.

\smallskip
\noindent
\textbf{Filter attack}~\cite{liu2019abs} applies Instagram filters on training samples and changes the ground truth labels of these samples to the target class. There are various filters can be used to poison data, such as Gotham filter, Nashville filter, Kelvin filter, Lomo filter, Toaster filter, etc.

\begin{table}[t]
    \caption{Comparison of different methods on hardening class distance for adversarially trained models (cont.)}
    \label{tab:harden_adv_app}
    \scriptsize
    \centering
    \tabcolsep=2.8pt
    \begin{tabular}{cc*{8}{r}}
        \toprule
        D & M & Method & Acc. & Rob. & Time & Dist. & Increase & ADeg. & RDeg. \\

        \midrule
        \multirow{6}{*}{\rotatebox{90}{CIFAR-10}}
        & \multirow{6}{*}{\rotatebox{90}{ResNet50}}
        &   Adversarial & 78.45\% & 42.30\% & 973.47 & 32.98 & - & - & - \\

        & & NC          & 77.21\% & 42.70\% & 155.98 & 33.30 & 1.08\% & 1.24\% & 0.00\% \\

        & & NAD         & 77.60\% & 44.50\% & 6.11 & 34.28 & 3.92\% & 0.85\% & 0.00\% \\
        
        & & ANP         & 77.05\% & 33.70\% & 4.66 & 30.15 & -8.49\% & 1.40\% & 8.60\% \\

        & & \moth{}     & 77.63\% & 41.70\% & 91.90 & 35.05 & 6.51\% & 0.82\% & 0.60\% \\
        
        & & \ours{}     & 77.20\% & 42.70\% & 41.67 & \textbf{39.81} & \textbf{21.70\%} & 1.25\% & 0.00\% \\

        \midrule
        \multirow{12}{*}[-0.03in]{\rotatebox{90}{LISA}}
        & \multirow{6}{*}{\rotatebox{90}{CNN}}
        &   Adversarial & 75.43\% & 25.00\% & 5.00 & 97.82 & - & - & - \\

        & & NC          & 73.44\% & 24.43\% & 8.54 & 87.82 & -9.53\% & 1.99\% & 0.57\% \\

        & & NAD         & 70.17\% & 23.57\% & 0.14 & 98.26 & -2.02\% & 5.26\% & 1.43\% \\
        
        & & ANP         & 73.86\% & 23.00\% & 0.07 & 86.14 & -12.33\% & 1.57\% & 2.00\% \\

        & & \moth{}     & 74.72\% & 31.14\% & 17.47 & 95.72 & -3.57\% & 0.71\% & 0.00\% \\
        
        & & \ours{}     & 74.29\% & 30.86\% & 5.78 & \textbf{105.47} & \textbf{15.30\%} & 1.14\% & 0.00\% \\

        \cmidrule{2-10}
        & \multirow{6}{*}{\rotatebox{90}{ResNet20}}
        &   Adversarial & 80.54\% & 36.57\% & 17.40 & 92.81 & - & - & - \\

        & & NC          & 78.13\% & 41.86\% & 79.76 & 93.79 & 7.37\% & 2.41\% & 0.00\% \\

        & & NAD         & 80.97\% & 40.00\% & 0.93 & 88.73 & -1.08\% & 0.00\% & 0.00\% \\
        
        & & ANP         & 80.26\% & 20.71\% & 1.18 & 81.04 & -8.01\% & 0.28\% & 15.86\% \\

        & & \moth{}     & 81.25\% & 41.00\% & 43.34 & 77.27 & -19.89\% & 0.00\% & 0.00\% \\
        
        & & \ours{}     & 81.82\% & 36.71\% & 6.18 & \textbf{101.55} & \textbf{17.80\%} & 0.00\% & 0.00\% \\

        \midrule
        \multirow{6}{*}{\rotatebox{90}{GTSRB}}
        & \multirow{6}{*}{\rotatebox{90}{NiN}}
        &   Adversarial & 90.96\% & 68.70\% & 54.98 & 50.02 & - & - & - \\

        & & NC          & 91.55\% & 63.60\% & 111.72 & 36.11 & -23.39\% & 0.00\% & 5.10\% \\

        & & NAD         & 88.12\% & 54.30\% & 0.50 & 36.49 & -20.58\% & 2.84\% & 14.40\% \\

        & & ANP         & 89.67\% & 61.30\% & 0.54 & 37.11 & -21.09\% & 1.29\% & 7.40\% \\

        & & \moth{}     & 90.32\% & 70.50\% & 71.29 & 48.54 & 2.00\% & 0.64\% & 0.00\% \\
        
        & & \ours{}     & 90.37\% & 69.20\% & 27.83 & \textbf{52.14} & \textbf{11.73\%} & 0.59\% & 0.00\% \\
        
        \midrule
        \multicolumn{2}{c}{\multirow{6}{*}{Average}}
        &   Adversarial & 81.35\% & 43.14\% & 262.71 & 68.41 & - & - & - \\

        & & NC          & 80.08\% & 43.15\% & 89.00 & 62.75 & -6.12\% & 1.27\% & 0.00\% \\

        & & NAD         & 79.22\% & 40.59\% & 1.92 & 64.44 & -4.94\% & 2.13\% & 2.55\% \\

        & & ANP         & 80.21\% & 34.68\% & 1.61 & 58.61 & -12.48\% & 1.13\% & 8.46\% \\

        & & \moth{}     & 80.98\% & 46.09\% & 56.00 & 64.15 & -3.74\% & 0.37\% & 0.00\% \\

        & & \ours{}     & 80.92\% & 44.87\% & 20.37 & \textbf{74.74} & \textbf{16.63\%} & 0.42\% & 0.00\% \\

        \bottomrule
    \end{tabular}
\end{table}

\subsection{Metrics}
\label{app:metrics}
For measuring the distance $d_{i \rightarrow j}$, we select 100 random samples from the validation set of a victim class $i$. Our backdoor generation is then applied for 300 epochs to generate a backdoor that can induce misclassification to the target class $j$ for 90\% of those samples. We carry out the above evaluation process for three times using different random seeds to alleviate potential influence from random initialization, different sets of samples, etc. For TrojAI models, it is quite slow to measure the distance for every class pair (taking days to evaluate one single model). We hence randomly select 20 class pairs for evaluating each model. To avoid selecting the same set of class pairs for models with the same number of classes, we set the random seed to be the sum of $165838010$ and the model id. The smallest distance obtained from the three trails is regarded as the final class distance.

\begin{table}[!ht]
    \caption{Comparison of different methods on hardening class distance for TrojAI models}
    \label{tab:harden_trojai}
    \scriptsize
    \centering
    \tabcolsep=4.7pt
    \begin{tabular}{ccc*{6}{r}}
        \toprule
        D & M & ID & Method & Accuracy & Time (m) & Distance & Increase & Degrad. \\

        \midrule
        \multirow{5}{*}[0.03in]{\rotatebox{90}{Swedish Roads}}

        & \multirow{5}{*}{\rotatebox{90}{ResNet101}}
        & \multirow{5}{*}{\rotatebox{90}{1}}
        & Natural       & 99.82\% & 1202.96 & 4.69 & - & - \\

        & & & NAD       & 99.67\% & 20.21 & 3.53 & -4.97\% & 0.15\% \\

        & & & ANP       & 99.81\% & 10.24 & 4.04 & -12.42\% & 0.01\% \\

        & & & \moth{}   & 99.79\% & 186.09 & 4.61 & 27.76\% & 0.04\% \\
        
        & & & \ours{}   & 99.67\% & 143.52 & 13.10 & 259.66\% & 0.15\% \\

        \midrule
        \multirow{15}{*}[-0.05in]{\rotatebox{90}{KITTI City}}

        & \multirow{5}{*}{\rotatebox{90}{VGG11bn}}
        & \multirow{5}{*}{\rotatebox{90}{111}}
        & Natural       & 99.86\% & 582.55 & 3.42 & - & - \\

        & & & NAD       & 99.83\% & 21.26 & 3.80 & 11.53\% & 0.03\% \\

        & & & ANP       & 99.78\% & 5.24 & 2.78 & -4.91\% & 0.08\% \\

        & & & \moth{}   & 99.45\% & 97.16 & 3.58 & 14.88\% & 0.42\% \\
        
        & & & \ours{}   & 99.65\% & 77.38 & 3.82 & 31.14\% & 0.22\% \\

        \cmidrule{2-9}
        & \multirow{5}{*}{\rotatebox{90}{ResNet34}}
        & \multirow{5}{*}{\rotatebox{90}{567}}
        & Natural       & 99.88\% & 339.35 & 4.37 & - & - \\

        & & & NAD       & 99.64\% & 9.18 & 2.80 & -12.58\% & 0.25\% \\

        & & & ANP       & 99.77\% & 3.98 & 3.38 & 6.92\% & 0.11\% \\

        & & & \moth{}   & 99.77\% & 62.86 & 3.96 & 17.10\% & 0.12\% \\
        
        & & & \ours{}   & 99.83\% & 87.60 & 8.58 & 80.78\% & 0.06\% \\

        \cmidrule{2-9}
        & \multirow{5}{*}{\rotatebox{90}{SqueezeNet1.0}}
        & \multirow{5}{*}{\rotatebox{90}{730}}
        & Natural       & 99.94\% & 205.43 & 3.13 & - & - \\

        & & & NAD       & 99.83\% & 7.02 & 3.06 & 1.59\% & 0.11\% \\

        & & & ANP       & 99.87\% & 3.60 & 2.74 & -1.70\% & 0.06\% \\

        & & & \moth{}   & 99.76\% & 38.36 & 3.08 & 13.88\% & 0.17\% \\
        
        & & & \ours{}   & 99.87\% & 67.41 & 4.08 & 53.59\% & 0.06\% \\

        \midrule
        \multirow{10}{*}[-0.05in]{\rotatebox{90}{KITTI Residential}}

        & \multirow{5}{*}{\rotatebox{90}{InceptionV3}}
        & \multirow{5}{*}{\rotatebox{90}{692}}
        & Natural       & 99.84\% & 545.91 & 5.37 & - & - \\

        & & & NAD       & 99.66\% & 12.32 & 5.75 & 17.74\% & 0.18\% \\

        & & & ANP       & 99.30\% & 7.01 & 4.79 & 9.93\% & 0.54\% \\

        & & & \moth{}   & 99.73\% & 282.45 & 4.50 & 16.20\% & 0.11\% \\
        
        & & & \ours{}   & 99.82\% & 78.87 & 5.77 & 48.05\% & 0.01\% \\

        \cmidrule{2-9}
        & \multirow{5}{*}{\rotatebox{90}{SqueezeNet1.1}}
        & \multirow{5}{*}{\rotatebox{90}{810}}
        & Natural       & 99.88\% & 253.15 & 4.26 & - & - \\

        & & & NAD       & 99.72\% & 9.86 & 4.74 & 29.82\% & 0.15\% \\

        & & & ANP       & 99.88\% & 4.16 & 4.41 & 18.93\% & 0.00\% \\

        & & & \moth{}   & 99.88\% & 33.99 & 4.91 & 35.78\% & 0.00\% \\
        
        & & & \ours{}   & 99.91\% & 177.18 & 6.34 & 57.15\% & 0.00\% \\

        \midrule
        \multirow{10}{*}[-0.05in]{\rotatebox{90}{KITTI Road}}

        & \multirow{5}{*}{\rotatebox{90}{ResNet18}}
        & \multirow{5}{*}{\rotatebox{90}{363}}
        & Natural       & 99.85\% & 561.59 & 4.10 & - & - \\

        & & & NAD       & 99.42\% & 14.10 & 5.01 & 21.25\% & 0.43\% \\

        & & & ANP       & 99.76\% & 5.36 & 5.03 & 13.88\% & 0.09\% \\

        & & & \moth{}   & 99.57\% & 40.47 & 7.74 & 86.73\% & 0.28\% \\
        
        & & & \ours{}   & 99.78\% & 144.72 & 9.92 & 104.92\% & 0.07\% \\

        \cmidrule{2-9}
        & \multirow{5}{*}{\rotatebox{90}{DenseNet121}}
        & \multirow{5}{*}{\rotatebox{90}{781}}
        & Natural       & 99.92\% & 654.97 & 2.93 & - & - \\

        & & & NAD       & 99.74\% & 17.30 & 2.44 & -9.68\% & 0.18\% \\

        & & & ANP       & 99.94\% & 10.96 & 3.05 & 7.07\% & 0.00\% \\

        & & & \moth{}   & 99.74\% & 121.62 & 3.31 & 22.98\% & 0.19\% \\
        
        & & & \ours{}   & 99.92\% & 98.31 & 4.10 & 40.47\% & 0.00\% \\

        \midrule
        \multirow{10}{*}[-0.05in]{\rotatebox{90}{Cityscapes}}

        & \multirow{5}{*}{\rotatebox{90}{ShuffleNet1.0}}
        & \multirow{5}{*}{\rotatebox{90}{210}}
        & Natural       & 99.96\% & 507.96 & 3.18 & - & - \\

        & & & NAD       & 99.95\% & 7.70 & 3.55 & 22.16\% & 0.01\% \\

        & & & ANP       & 99.94\% & 3.57 & 3.27 & 6.19\% & 0.02\% \\

        & & & \moth{}   & 99.92\% & 50.48 & 4.65 & 70.05\% & 0.04\% \\
        
        & & & \ours{}   & 99.96\% & 170.38 & 5.73 & 88.17\% & 0.00\% \\

        \cmidrule{2-9}
        & \multirow{5}{*}{\rotatebox{90}{SqueezeNet1.0}}
        & \multirow{5}{*}{\rotatebox{90}{621}}
        & Natural       & 99.93\% & 231.30 & 4.67 & - & - \\

        & & & NAD       & 99.92\% & 7.37 & 3.76 & -2.35\% & 0.02\% \\

        & & & ANP       & 99.94\% & 3.14 & 3.31 & -7.25\% & 0.00\% \\

        & & & \moth{}   & 99.82\% & 83.70 & 6.58 & 38.83\% & 0.12\% \\
        
        & & & \ours{}   & 99.91\% & 78.57 & 7.83 & 124.15\% & 0.02\% \\

        \midrule
        \multicolumn{3}{c}{\multirow{6}{*}{Average}}
        &   Natural     & 99.89\% & 508.52 & 4.01 & - & - \\

        & & & NAD       & 99.74\% & 12.63 & 3.84 & 7.45\% & 0.15\%\\

        & & & ANP       & 99.80\% & 5.73 & 3.68 & 3.66\% & 0.09\% \\

        & & & \moth{}   & 99.74\% & 99.72 & 4.69 & 34.42\% & 0.15\% \\

        & & & \ours{}   & 99.83\% & 112.39 & 6.93 & 88.81\% & 0.06\% \\
        \bottomrule
    \end{tabular}
\end{table}

\section{Additional Results on Enlarging Class Distance}
\label{app:harden_distance}

From~\autoref{tab:harden_nat_app}, observe that \ours{} can enlarge the class distance from 31.16 to 48.26 on average (57.37\% improvement), with 0.31\% and 0.35\% degradation on accuracy and robustness, respectively. \ours{} has the largest distance improvement on CNN with LISA, with 111.38\% enlargement. The three backdoor removal techniques, NC, NAD, and ANP, have limited performance on the class distance, with less than 8\% improvement on average, which is consistent with the observation on CIFAR-10 and SVHN datasets and discussed in Section~\ref{sec:eval_harden}. NC takes 84.05 minutes on average to harden a model, even longer than training the original model (20.14 minutes). NAD and ANP have low time costs, with 1.85 and 1.58 minutes overhead respectively. UAP can increase the class distance for 7.44\% on NiN with GTSRB. But its performance on other datasets and models are still limited, with an average of 3.20\% enlargement. In addition, UAP is extremely slow, taking 93.57 minutes on average, which is 4.6x longer than the natural training. \moth{} can improve the class distance by more than 10\% in one cases (CNN on LISA). However, \moth{} cannot enlarge the class distance for ResNet20 on LISA. The time cost of \moth{} is twice of training a natural model, with 44.39 minutes overhead. Overall, \ours{} has the largest class distance improvement for the evaluated models, substantially outperforming baselines. \ours{} has a reasonable hardening cost, with 18.86 minutes overhead on average.

Similar observations can also be made on adversarially trained models in~\autoref{tab:harden_adv_app}. \ours{} is able to enlarge the class distance by 16.63\% on average, with only 0.42\% accuracy degradation and no robustness loss. NC, NAD, and ANP have worse performance on adversarially trained models than on naturally trained ones, with -6.12\%, -4.94\%, and -12.48\% enlargement, respectively. In addition, NAD and ANP have more than 2.5\% robustness degradation, which is critical to adversarially trained models. \moth{} has a reasonable performance on CIFAR-10, improving the class distance by 6.51\%. The results on LISA and GTSRB are inferior. The hardening time costs of all the methods are similar to those on naturally trained models.

\section{Comparison with Neural Collapse}
\label{app:neuron_collapse}

Existing work~\cite{papyan2020prevalence} observes that by training the model to the terminal phase (training beyond zero misclassification error), the model can exhibit better generalization performance, better robustness, etc., which is called Neural Collapse.
We follow the existing work~\cite{papyan2020prevalence} by training a ResNet20 model on CIFAR-10 to 250 epochs and 300 epochs. The class distance for the two models are 24.46 and 24.63, respectively, slightly larger than the original model (22.62). The model hardened by \ours{} has a distance of 34.60. This is because the normal training does not aim to enlarge pairwise class distances but only reduces prediction loss. More training epochs can hardly improve the distance.

\section{Correlation of Distance Measures by \moth{} and \ours{}}
\label{app:distance_correlation}

We use the distance in \moth{} to measure a hardened ResNet20 model on CIFAR-10 by \ours{}. The distance value increases from 53.49 to 64.29, whereas \moth{} can increase the distance to 109.70. This shows that \ours{} can increase the \moth{} distance to some extent. The distance measured in our paper increases from 22.62 to 25.92 for MOTH, to 34.60 for \ours{}. The two measures focus on different aspects of class distance, analogous to $\normlzero$ and $\normltwo$ norms in adversarial robustness. Existing works have shown that adversarial robustness on one norm does not promise robustness on another norm~\cite{tramer2019adversarial}, which is consistent with the above observation in backdoor robustness.

\section{Comparison with More Baselines on Eliminating Injected Pervasive backdoors}
\label{app:removal}

\begin{table*}[t]
    \caption{Evaluation on backdoor attacks with more baselines}
    \label{tab:backdoors_app}
    \scriptsize
    \centering
    \begin{tabular}{cll*{8}{r}}
        \toprule
        \multirow{2}{*}[-0.03in]{\parbox{0.5in}{\centering Backdoor Attack}} & \multirow{2}{*}[-0.03in]{Dataset} & \multirow{2}{*}[-0.03in]{Model} & \multicolumn{2}{c}{Original} & \multicolumn{2}{c}{Fine-pruning} & \multicolumn{2}{c}{MCR} & \multicolumn{2}{c}{\ours{}} \\

        \cmidrule(lr){4-5} \cmidrule(lr){6-7} \cmidrule(lr){8-9} \cmidrule(lr){10-11}
        & & & Accuracy & ASR & Accuracy & ASR & Accuracy & ASR & Accuracy & ASR \\

        \midrule
        \multirow{8}{*}{DFST}
        & \multirow{4}{*}{CIFAR-10}
        & ResNet32 Detox1
        & 89.95\% & 97.60\% & 88.20\% & 62.89\% & 87.95\% & 62.67\% & 88.22\% & \underline{14.22\%} \\
        
        & & ResNet32 Detox3
        & 90.93\% & 95.33\% & 88.09\% & 47.44\% & 82.84\% & 60.78\% & 88.11\% & \underline{12.22\%} \\
        
        & & VGG13 Detox1
        & 90.34\% & 95.89\% & 87.37\% & 51.11\% & 86.58\% & 90.33\% & 88.03\% & \underline{2.00\%} \\
        
        & & VGG13 Detox3
        & 91.29\% & 97.44\% & 88.84\% & 85.67\% & 88.81\% & 86.78\% & 89.08\% & \underline{5.67\%} \\
        
        \cmidrule{2-11}
        
        & \multirow{4}{*}{STL-10}
        & ResNet32 Detox1
        & 75.74\% & 97.67\% & 68.89\% & 96.11\% & 68.05\% & 84.67\% & 72.10\% & \underline{2.67\%} \\
        
        & & ResNet32 Detox3
        & 76.45\% & 99.00\% & 69.25\% & 88.22\% & 69.98\% & 81.89\% & 72.86\% & \underline{4.78\%} \\
        
        & & VGG13 Detox1
        & 72.18\% & 98.67\% & 67.12\% & 67.00\% & 66.06\% & 66.22\% & 68.61\% & \underline{5.89\%} \\
        
        & & VGG13 Detox3
        & 72.09\% & 98.89\% & 68.14\% & 49.44\% & 66.66\% & 79.67\% & 69.89\% & \underline{12.33\%} \\
        
        \midrule
        \multirow{2}{*}{Blend}
        & CIFAR-10 & ResNet20
        & 90.96\% & 99.96\% & 87.75\% & 3.63\% & 85.53\% & 63.58\% & 89.08\% & \underline{0.00\%} \\
        
        & SVHN & NiN
        & 94.10\% & 92.37\% & 88.26\% & 23.75\% & 93.50\% & \underline{0.59\%} & 94.56\% & 0.85\% \\
        
        \midrule
        \multirow{2}{*}{SIG}
        & CIFAR-10 & ResNet20
        & 83.38\% & 93.30\% & 81.01\% & 76.29\% & 83.31\% & 16.63\% & 86.91\% & \underline{3.97\%} \\
        
        & SVHN & NiN
        & 95.48\% & 92.46\% & 93.35\% & 23.49\% & 93.19\% & 45.16\% & 93.96\% & \underline{0.46\%} \\

        \midrule
        \multirow{3}{*}{WaNet}
        & CIFAR-10 & ResNet18
        & 94.15\% & 99.55\% & 89.14\% & 2.09\% & 93.29\% & 1.74\% & 91.12\% & \underline{0.64\%} \\
        
        & GTSRB & ResNet18
        & 99.01\% & 98.94\% & 96.06\% & 63.40\% & 98.54\% & 10.47\% & 97.70\% & \underline{0.30\%} \\
        
        & CelebA & ResNet18
        & 78.99\% & 99.08\% & 76.57\% & 18.07\% & 78.32\% & 16.21\% & 77.57\% & \underline{8.12\%} \\
        
        \midrule
        Clean Label & CIFAR-10 & ResNet18
        & 87.60\% & 98.36\% & 83.66\% & 26.77\% & 85.23\% & 13.91\% & 85.67\% & \underline{4.22\%} \\
        
        \midrule
        
        \multicolumn{3}{c}{\textbf{Average}}
        & \textbf{86.42\%} & \textbf{97.16\%} & \textbf{82.61\%} & \textbf{49.09\%} & \textbf{82.99\%} & \textbf{48.83\%} & \textbf{84.59\%} & \textbf{\underline{4.90\%}} \\

        \bottomrule
    \end{tabular}
\end{table*}

The results of two more baselines, namely, Fine-pruning and MCR on eliminating injected pervasive backdoors are shown in~\autoref{tab:backdoors_app}. Columns 6-7 present the accuracy and ASR of models repaired by Fine-pruning and columns 8-9 are for MCR. We also show the results of the original models and models repaired by \ours{} in columns 4-5 and columns 10-11 respectively for comparison. It is evident that both Fine-pruning and MCR have limited defense performance against DFST with only around 50\% ASR reduction at best. As discussed in Section~\ref{sec:eval_removal}, DFST leverages a detoxification procedure to make injected backdoors more robust against defenses. Fine-pruning is not able to identify compromised neurons and hence less effective. Since Standard Finetune fails to remove DFST-injected backdoors (see~\autoref{tab:backdoors} in Section~\ref{sec:eval_removal}), there does not exist a backdoor-free model along the path explored by MCR from the poisoned model to the finetuned version. Fine-pruning can eliminate backdoors for a few models poisoned by Blend and WaNet (rows 11 and 15). However, it still fails to completely remove injected backdoors for the remaining cases. The observation is similar for MCR with two cases of the ASR lower than 5\%. Overall, Fine-pruning/MCR can reduce the ASR to 49.09\%/48.83\% with 3.81\%/3.43\% accuracy degradation on average. In contrast, \ours{} has the ASR reduction from 97.16\% to 4.90\% and the accuracy degradation is only 1.83\%, significantly outperforming the two baselines.

\section{Ablation Study}
\label{app:ablation}

\begin{table}[t]
    \caption{Ablation study on effects of different design choices}
    \label{tab:ablation}
    \scriptsize
    \centering
    \tabcolsep=4.2pt
    \begin{tabular}{l*{6}{r}}
        \toprule
        Method & Accuracy & Time (min) & Distance & Increase & Difference \\

        \midrule
        Natural     & 91.52\% & 56.77 &  22.62 & - & - \\

        \ours{}     & 90.31\% & 16.73 & 34.60 & 52.66\% & - \\
        
        \cmidrule(lr){1-6}
        w/ SWWAE~\cite{zhao2015stacked,zhang2016augmenting}    & 90.97\% & 16.48 & 23.75 &  4.63\% & -48.03\% \\
        w/ Nearest~\cite{odena2016deconvolution}  & 90.32\% & 16.37 & 23.35 &  3.52\% & -49.14\% \\
        w/ Zero~\cite{dosovitskiy2016inverting}     & 89.83\% & 16.55 & 30.92 & 36.30\% & -16.36\% \\
        w/ Pool2    & 90.81\% & 19.61 & 24.93 & 10.34\% & -42.32\% \\

        \cmidrule(lr){1-6}
        w/o $\mathcal{L}_{content}$ & 90.13\% & 16.47 & 33.42 & 46.48\% & -6.18\% \\
        w/o $\mathcal{L}_{SSIM}$    & 90.36\% & 17.25 & 29.15 & 27.91\% & -24.75\% \\
        w/o $\mathcal{L}_{norm}$    & 90.27\% & 17.88 & 30.10 & 32.38\% & -20.28\% \\
        w/o $\mathcal{L}_{smooth}$  & 90.13\% & 18.09 & 33.64 & 47.54\% & -5.12\% \\

        \bottomrule
    \end{tabular}
\end{table}
\ours{} features a few important design choices. In this section, we aim to study these choices individually to better understand their contributions to the performance. In particular, we first evaluate on different decoder designs as well as the position of the feature representation in the encoder. We then study the effects of four loss terms used in pervasive backdoor generation. The ablation study is conducted on a ResNet20 model on CIFAR-10 and the results are presented in~\autoref{tab:ablation}. Row 2 denotes the original model and row 3 the final result of \ours{}. Rows 4-6 show the results of using different decoders. Row 7 denotes using the Pool2 layer in the encoder as the feature representation. Rows 8-11 present the results of excluding each loss term individually during hardening. Observe that using SWWAE has only 4.63\% class distance improvement. This is because the feature representation of backdoor samples can be very different from normal samples. The positional information of normal samples in the encoder leveraged by SWWAE does not help decoding the transformed representation that are similar to that of backdoor samples. Using Nearest has only 3.52\% improvement and hardening using Zero leads to 16.36\% improvement degradation. Nearest and Zero decoders impose a strong linear correlation among upsampled values, which reduces the quality of generated backdoor samples. The default setting of \ours{} employs the Pool1 layer of the encoder for extracting the feature representation, which is the first pooling layer in the encoder. Here, we study using a deeper layer for feature extraction. The result in row 7 shows that using a deeper layer reduces the quality of decoded samples as more errors are introduced during upsampling in the decoder and causes a smaller class distance improvement. For the four loss terms, we observe that $\gL_{SSIM}$ has the largest impact on the final result (24.75\% degradation without it). This is because the SSIM score directly constrains the quality of generated backdoor samples looking similar to original inputs. Without it, the backdoor samples can be too different from normal inputs and the model cannot learn the correct features. Loss $\gL_{norm}$ on the normalization layer is also quite important (20.28\% degradation without it) as it makes sure the normalized inputs are not far from the original distribution. Loss $\gL_{content}$ constrains the difference of feature representations between backdoor samples and normal inputs. The purpose is similar to the SSIM loss and it can boost the performance by 6.18\%. Loss $\gL_{smooth}$ smooths the backdoor samples to improve the generated quality, which can provide 5.12\% performance gain.

\end{document}